\documentclass[12pt]{article}
\usepackage{amssymb,epsfig,times,graphicx}
\usepackage{amsmath}
\usepackage{pstricks,pst-node,pst-plot}
\usepackage{graphicx}
\numberwithin{equation}{section}

\newcommand{\im} {\mbox {Im}\hskip 0.5truemm}

\newtheorem{thm}{Theorem}

\newtheorem{prop}{Proposition}

\newtheorem{lemma}{Lemma}

\newtheorem{cor}{Corollary}

\newtheorem{exam}{Example}

\newtheorem{rmk}{Remark}

\newcommand{\eqa}{\begin{eqnarray}}
\newcommand{\eeqa}{\end{eqnarray}}
\newcommand{\beq}{\begin{equation}}
\newcommand{\eeq}{\end{equation}}

\newcommand{\de}{\partial}

\newcommand{\la}{\lambda}

\newcommand{\pf}{\noindent{\it Proof \ }}

\newcommand{\epf}{$\quad$\hfill
\raisebox{0.11truecm}{\fbox{}}\par\vskip0.4truecm}

\newcommand{\sch}{\mathcal{S}(\mathbb{R})} 
\newcommand{\tem}{\mathcal{S}'(\mathbb{R})}
\newcommand{\schd}{\mathcal{S}(\mathbb{R}^{2})}
\newcommand{\temd}{\mathcal{S}'(\mathbb{R}^{2})} 
\newcommand{\om}{\mathcal{O}_M(\mathbb{R})}
\newcommand{\omd}{\mathcal{O}_M(\mathbb{R}^{2})}

\def\Xint#1{\mathchoice
   {\XXint\displaystyle\textstyle{#1}}%
   {\XXint\textstyle\scriptstyle{#1}}%
   {\XXint\scriptstyle\scriptscriptstyle{#1}}%
   {\XXint\scriptscriptstyle\scriptscriptstyle{#1}}%
   \!\int}
\def\XXint#1#2#3{{\setbox0=\hbox{$#1{#2#3}{\int}$}
     \vcenter{\hbox{$#2#3$}}\kern-.5\wd0}}

\def\dashint{\Xint-}

\begin{document}
 
\title{Frobenius manifold for the dispersionless Kadomtsev-Petviashvili equation}
\author{Andrea Raimondo\\
\\
{\small SISSA}\\
{\small Via Bonomea 265, 34136 Trieste, Italy}\\
{\small andrea.raimondo@sissa.it}\\
}
\date{}
\maketitle

\begin{abstract}
We consider a Frobenius structure associated with the dispersionless Kadomtsev -- Petviashvili equation. This is done, essentially, by applying a continuous analogue of the finite dimensional theory in the space of Schwartz functions on the line. The potential of the Frobenius manifold is found to be a logarithmic potential with quadratic external field. Following the construction of the principal hierarchy, we construct a set of infinitely many commuting flows, which extends the classical dKP hierarchy.
\end{abstract}

\section*{Introduction}
The relation between Frobenius manifolds and hierarchies of integrable dispersionless system, first described by Dubrovin \cite{du92,du96}, has been subject in the last twenty years of an intense research (see \cite{duzh01,duzh06} and references therein). The connection is obtained by considering a class of integrable quasilinear $(1\!+\!1)$ PDEs, of the form
$$u^{i}_{t}=V^{i}_{j}(u)\,u^{j}_{x},$$ 
which are known as hydrodynamic type systems. To every Frobenius manifold, one can associate a hierarchy of infinitely many commuting systems of hydrodynamic type, whose number of components is equal to the dimension of the manifold. This hierarchy goes under the name of principal hierarchy, and the explicit knowledge of the members of the hierarchy provides the general solution of any of its flows, by a procedure known as generalized hodograph method. In this setting, the knowledge of a Frobenius manifold is therefore a useful tool for the integration of these systems. The converse problem, namely to find a Frobenius manifold starting from a system of hydrodynamic type, is rather more difficult, for one has to provide a suitable decomposition of the matrix $V$ in terms of structure constants of an algebra. If one is able to specify the Frobenius manifold, then the flows of the corresponding principal hierarchy provide a complete set of symmetries for the original system, and the hodograph method can be applied. In this paper, we are interested in infinite dimensional Frobenius manifolds, to be associated with $(2\!+\!1)-$equations.

The first complete example of an infinite dimensional Frobenius manifold has been provided by Carlet, Dubrovin and Mertens in the recent paper \cite{cadume09}, where they have related the dispersionless $2D$ Toda hierarchy to a Frobenius manifold constructed on the space of pairs of functions analytic inside and outside the unit circle respectively, and with prescribed singularities at zero and at infinity. The purpose of the present paper is to produce a Frobenius structure for the pair of commuting flows
\begin{subequations}\label{vlasovintro}
\begin{align}
f_y&=pf_x-A^0_xf_p,\\
f_t&=(p^{2}+A^{0})f_x-(A^0_x\,p+A^{1}_{x})f_p,
\end{align}
\end{subequations}
which are two examples of kinetic equations of Vlasov type. Here $A^k\!=\!\int_{-\infty}^{+\infty} p^{k}f\,dp$ are the moments of $f$ with respect to the variable $p$, and subscripts denote partial derivatives. The above pair of equations is strictly related with the $(2\!+\!1)-$equation
\beq\label{dkp}
\de_x\left(A^{0}_t-A^{0}A^{0}_x\right)=A^{0}_{yy},
\eeq
which is known in the literature as \emph{dispersionless Kadomtsev--Petviashvili} (dKP) equation. More precisely, the first moment $A^{0}$ satisfies \eqref{dkp} provided $f$ is a solution of \eqref{vlasovintro}. Equation \eqref{dkp} can be derived as the dispersionless limit of the Kadomtsev--Petviashvili equation \cite{kp70}, and it also appears in nonlinear acoustics under the name of Khokhlov--Zabolotskaya equation \cite{zakh69,ruso77}.

Despite the fact that the dKP and the dispersionless $2D$ Toda equations present many similarities, the procedure established in \cite{cadume09} for constructing a Frobenius manifold seems to be less suitable, as it stands,  for the dKP case. The present approach is thus different, and based on a method introduced in \cite{gibrai07} for equations of type \eqref{vlasovintro}. The crucial point is to interpret these systems as a sort of systems of hydrodynamic type, of the form
$$f_{t}(p)=\int V\tbinom{p}{q}f_{x}(q)dq,$$
where $V$ is a suitable kernel, and to construct from these systems a Frobenius structure by applying a direct generalization of the finite dimensional case.  There are two important features we want to emphasize: first,  although the classical theory of Frobenius manifolds is usually easier described in the so-called flat coordinates, we prefer in this paper to use a different set of coordinates -- given essentially by the function $f(p)$ -- which turns out to be more suitable for our purpose. In addition, in order to give precise sense to the objects involved in the Frobenius structure, we  assume $f$ to be a Schwartz function of the variables $x$ and $p$. Correspondingly, the Frobenius structure is thus provided by considering (multi)linear maps on suitable vector subspaces of the Schwartz space and its dual space, the space of tempered distributions.

The choice of taking $f$ in the Schwartz class implies that the first moment $A^{0}$ -- which is the solution of the dKP equation -- is a Schwartz function in the variable $x$. Within this approach, the behaviour of $A^{0}$ for large $y$ is left arbitrary. In this paper, we do not consider the Cauchy problem for the dKP equation (see \cite{masa06,masa07,masa08}, where a new kind of inverse scattering transform has been introduced for solving the dKP equation with localized initial data); however -- following the construction of the principal hierarchy of the Frobenius manifold -- we produce a hodograph-type formula, which provides solutions of dKP in an implicit form. 

The paper is organized as follows: in Section \ref{secfrob} we review the finite dimensional theory of Frobenius manifolds, and the relation with hydrodynamic type systems. Section \ref{secvlbe} is devoted to the Lax representation of the dKP hierarchy, to its Hamiltonian formulation, and to the kinetic representation of its flows, which can be written as Vlasov equations. In Section \ref{section:sch}, we set up the analytical background, we describe the space of Schwartz functions (on the line and on the plane), and the corresponding dual space of tempered distributions. Furthermore, we review the classical results concerning Schwartz functions, their Hilbert transform, and scalar Riemann--Hilbert problem on the real axis. 

Section \ref{sec:from} is devoted to the construction of the Frobenius manifold. Using the continuous-index approach, we first show how to write systems of type \eqref{vlasovintro} as hydrodynamic type systems, we rewrite in this picture the metrics associated of the dKP Poisson brackets,  and we  determine suitable structure constants, the unity and Euler vector fields, and the potential of the Frobenius manifold. The latter turns out to be a logarithmic potential with external field, of the form
$$F=\frac{1}{2}\iint \log{|p-q|}f(p)f(q)dp\,dq+\frac{1}{2}\int p^2f(p)dp.$$

In Section \ref{secprinchie}, we follow the procedure of the principal hierarchy of a Frobenius manifold, and we construct an infinite set of commuting flows, which generalizes the classical dKP hierarchy. We then introduce an analogue of the hodograph transform, and discuss the validity of this formulation in the infinite dimensional setting.

As already mentioned, the construction of the Frobenius manifold is done in this paper without using flat coordinates. For completeness, in the last section we describe the flat coordinates for the Frobenius manifold, as well as the canonical coordinates, which diagonalize the flows of the principal hierarchy. The construction of these coordinate sets corresponds to the procedure followed by Carlet, Dubrovin and Mertens in the case of the dispersionless $2D$ Toda hierarchy.

\section{Finite dimensional Frobenius manifolds}\label{secfrob}
In this section we briefly review the theory of Frobenius manifolds, as well as their connection with integrable hierarchies. The main purpose here is to fix the notation and to explicitly write down the formulae we want to generalize later to the infinite dimensional case.
For a complete exposition on Frobenius manifolds and its related topics we refer to \cite{du96,duzh01}.

\subsection{Frobenius manifolds}
A \emph{Frobenius algebra} is a pair $(A,<\,\,,\,>)$, where $A$ is a commutative, associative algebra with unity over a field $k$ ($=\mathbb{R},\mathbb{C}$), and $<\,\,,\,>$ is a $k-$bilinear, symmetric, non-degenerate form, which is required to be invariant, in the sense that
$$<X\circ Y,Z>=<X,Y\circ Z>, \qquad \forall\,\, X,Y,Z\in A.$$
A \emph{Frobenius structure} of charge $d$ on a differentiable manifold $\mathcal{M}$ is the structure of a Frobenius algebra on every tangent space $T_z\mathcal{M}$, depending smoothly on $z\in \mathcal{M}$. Moreover, for every vector field $X$, $Y$, $Z$, $W$ on $\mathcal{M}$, the following conditions are required:
\begin{enumerate}
\item The pseudo-metric
$$\eta\left(X,Y\right):=<X,Y>$$
on $\mathcal{M}$ is flat.
\item Denoting by $\nabla$ the Levi-Civita connection of the metric $\eta$, then the (totally symmetric) tensor
$$c\left(X,Y,Z\right):=\eta\left(X\circ Y,Z\right),$$
must satisfy
$$\nabla_{\!W}c\left(X,Y,Z\right)=\nabla_Xc\left(W,Y,Z\right).$$
\item The unity vector field $e$ is flat:
$$\nabla e=0.$$
\item There exists a vector field $E$ on $\mathcal{M}$, called \emph{Euler} vector field, satisfying
\begin{align*}
&\nabla \nabla E=0,\\
&\left[E,X\circ Y\right]-\left[E,X\right]\circ Y-X\circ\left[E,Y\right]= X\circ Y.\\
&E\left(\eta\left(X,Y\right)\right)-\eta\left(\left[E,X\right],Y\right)-\eta\left(X,\left[E,Y\right]\right)=(2-d)\,\eta\left(X,Y\right).
\end{align*}
\end{enumerate}
The above definition gives to a Frobenius manifold a very rich geometrical structure, which has been object of intense studies in the last twenty years \cite{du96,duzh01}. In particular, one of the most important results is the existence of a function $F$, called the \emph{potential} of the Frobenius manifold, satisfying, in any local coordinate set $u^1,\dots,u^N$, the conditions 
\beq\label{potanycoo}
c_{ijk}(u)=\nabla_i\nabla_j\nabla_k F(u).
\eeq
\begin{rmk}
By using flat coordinates for the metric $\eta$, say $h^1,\dots,h^N$, condition \eqref{potanycoo} reduces to
$$c_{ijk}(h)=\frac{\de^3 F(h)}{\de h^i\de h^j\de h^k},$$
and the associativity equations for the algebra can be written, in these coordinates, as
$$\frac{\de^3 F(h)}{\de h^i\de h^j\de h^s}\,\eta^{sr}\!\frac{\de^3 F(h)}{\de h^r\de h^l\de h^k}=\frac{\de^3 F(h)}{\de h^k\de h^j\de h^s}\,\eta^{sr}\!\frac{\de^3 F(h)}{\de h^r\de h^l\de h^i}.$$
These equations are known as WDVV equations, and play an important role in two dimensional topological field theory \cite{wit91,dvv91}
\end{rmk}
We denote by $\eta^{ij}$ the components of the inverse metric $\eta^{-1}$, so that the quantities
$$c^i_{jk}=\eta^{is}c_{sjk},$$
are the structure constants of the algebra. Another interesting feature for a Frobenius manifold is that, in addition to the flat metric $\eta$,  it is possible to introduce a second flat metric, denoted by $g$, whose contravariant components are given by
$$g^{ij}=E^k\eta^{is}c^j_{ks},$$
where $E$ is the Euler vector field. This metric is usually called the \emph{intersection form} of the Frobenius manifold, its flateness can be deduced from the properties of $E$, $\eta$ and $c$, and it has the important property that every linear combination of the form
$$\eta_{ij}+\alpha\, g_{ij}, \qquad \alpha=const.,$$
defines a metric with zero curvature tensor. The metrics $\eta$ and $g$ are thus said to be \emph{compatible flat metrics}, or to form a \emph{flat pencil of metrics} \cite{du97}.

Finally, we recall the notion of semisimple Frobenius manifold. A point $z$ on a Frobenius manifold $\mathcal{M}$ is said to be \emph{semisimple} if the corresponding tangent space $T_z\mathcal{M}$ is semisimple (that is, it has no nilpotents). A Frobenius manifold $\mathcal{M}$ is said to be \emph{semisimple} if a generic point of $\mathcal{M}$ is semisimple. In a neighborhood of a semisimple point of a Frobenius manifold, one can introduce local coordinates $r^1, \dots, r^N$ such that
$$c^i_{jk}(r)=\delta^i_j\delta^i_k.$$
Such coordinates are called \emph{canonical coordinates}; it can be shown that in these coordinates the metric and the intersection form take diagonal form, namely
$$\eta_{ij}(r)=\eta_{ii}(r)\delta_{ij},\qquad g_{ij}(r)=g_{ii}(r)\delta_{ij},$$
for suitable functions $\eta_{ii}$, $g_{ii}$.

\subsection{Integrable hierarchies and Frobenius manifolds}
In $(1+1)$ dimension, \emph{systems of hydrodynamic type} are systems of quasilinear PDEs of the form
\beq\label{hts}
u^i_t=V^i_j(u)u^j_x, \qquad i=1,\dots,N,
\eeq
where $V^i_j$ are differentiable functions of the $u^i(x,t)$, and the independent variables $x$ and $t$ are real. We also require the boundary conditions
\beq\label{dboundary}
\lim_{|x|\mapsto\infty}u^i(x,t)=0,\qquad i=1,\dots,N,
\eeq
to hold. Within these hypotheses, if we consider $u^1,\dots,u^N$ as local coordinates on a differentiable manifold $\mathcal{M}$, we can interpret systems of type \eqref{hts} as dynamical systems on the loop space
$$L(\mathcal{M})=\left\{\gamma:\mathbb{R}\mapsto \mathcal{M}\,\,|\,\, \gamma(x)=\left(u^1(x),\dots,u^N(x)\right)\right\},$$
of smooth curves on the manifold $\mathcal{M}$.

A system of type \eqref{hts} is said to be \emph{diagonalizable} if there exist coordinates $r^1,\dots,r^N$ such that the system is diagonal, that is, $V^i_j(r)=v^i(r)\delta^i_j,$ or, equivalently,
\beq\label{diag}
r^i_t=v^ir^i_x.
\eeq
Such coordinates are called \emph{Riemann invariants}, and the $v^i$ \emph{characteristic velocities}. Due to a result of Tsarev \cite{ts90}, solutions of a system \eqref{diag} can be given in an implicit way by the Hodograph formula
\beq\label{hodo}
x+tv^i=w^i,
\eeq
where the $w^i$ are solutions of the linear system
\beq\label{htssymm}
\frac{\de w^i}{\de r^j}=\frac{\frac{\de v^i}{\de r^j}}{v^j-v^i}\left(w^j-w^i\right),\qquad i\neq j.
\eeq
This system provides the characteristic velocities of the symmetries of \eqref{diag}; hydrodynamic type systems for which \eqref{htssymm} is compatible are known in the literature as \emph{semi-Hamiltonian}. Every semi-Ha\-mil\-to\-nian systems possesses, besides the symmetries solutions of \eqref{htssymm}, infinitely many conserved quantities.

Let us now move to the relation between Frobenius manifolds and hydrodynamic type systems. This can be described by introducing a Poisson structure on the loop space $L(\mathcal{M})$. Thus, we let $u^{1},\dots,u^{n}$ depend on $x$, we impose the boundary conditions \eqref{dboundary}, and -- following Dubrovin and Novikov \cite{duno83} -- we define \emph{functionals of hydrodynamic type} to be functionals of the form
\beq\label{funht}
H[u]=\int_{-\infty}^{+\infty}h(u)dx,
\eeq
where $h$ is a differentiable function, depending on the $u^i$ but not on their $x-$derivatives. A \emph{Poisson bracket of hydrodynamic type} of two functionals $H=\int hdx$ and $K=\int k dx$ of type \eqref{funht} is defined as
\beq\label{dnpoisson}
\left\{H,K\right\}=\int\frac{\delta H}{\delta u^i(x)}\left(\eta^{ij}(u)\frac{d}{dx}+\Gamma^{ij}_k(u)u^k_x\right) \frac{\delta K}{\delta u^j(x)}dx,
\eeq  
where the $\eta^{ij}$ are the contravariant components of a flat metric on $\mathcal{M}$, and $\Gamma^{ij}_k=-\eta^{is}\Gamma^j_{sk}$, with $\Gamma^j_{sk}$ the Christoffel symbols of the metric $\eta$. For functionals of type \eqref{funht}, the variational derivative involved in the formula is given by
\beq\label{fvdientity}
\frac{\delta H}{\delta u^i(x)}=\frac{\de h}{\de u^i}.
\eeq
It is thus clear that Hamiltonians of type \eqref{funht} generate Hamilton equations
$$u^i_t=\left\{u^i,H\right\}=\left(\eta^{ik}\nabla_k\nabla_jh\right)u^j_x,$$
which are systems of hydrodynamic type. Since a Frobenius manifold admits two compatible flat metrics, the loop space of a Frobenius manifold admits a bihamiltonian structure, whose flows generated by Hamiltonians of the form \eqref{funht} are systems of hydrodynamic type.

An alternative approach for constructing a hydrodynamic type system starting from a Frobenius manifold is to consider a vector field $X$ on $\mathcal{M}$, and to associate to $X$ the hydrodynamic type system
\beq\label{htsprod}
u^i_t=\left(V_X\right)^i_ju^j_x,\qquad \left(V_X\right)^i_j=c^i_{jk}X^k,
\eeq
where $c$ is the product of the Frobenius manifold. By combining these two approaches, one can construct an infinite set of commuting flows, known as the \emph{principal hierarchy} of the Frobenius manifold \cite{duzh01}. This can be defined by means of the recursive relations
\begin{subequations}\label{fdprinchie}
\begin{align}
&\nabla_j X^i_{\alpha, 0}=0,\label{flatvf}\\
&\nabla_j X^i_{\alpha, n+1}=c^i_{jk}X^k_{\alpha,n},\qquad n=0,1,2,\dots,\label{princ}
\end{align}
\end{subequations}
where $\alpha=1,\dots,N$, and $\nabla$ is the covariant derivative of the metric $\eta$ (see for instance \cite{lopera09}, where the above construction has been considered for a more general class of systems). If $c$ and $\eta$ satisfy the conditions of a Frobenius manifold, then system \eqref{fdprinchie} is compatible, and the flows
\beq\label{princflo}
\frac{\de u^i}{\de t_{\alpha,n}}=\left(V_{\alpha,n}\right)^i_j\frac{\de u^j}{\de x}, \qquad \left(V_{\alpha,n}\right)^i_j=c^i_{jk}X^k_{\alpha,n},
\eeq
which form the principal hierarchy, pairwise commute. In particular, when the flat vector fields $X_{\alpha, 0}$ are chosen as in \cite{duzh01}, the corresponding PDEs are called \emph{primary flows} of the hierarchy.  The flows of the principal hierarchy are semi-Hamiltonian hydrodynamic type systems, and therefore they possess infinitely many conserved quantities of hydrodynamic type, whose densities are solutions of the system
\beq\label{consdenhie}
c^i_{jk}\nabla_i\nabla_l h=c^i_{lk}\nabla_i\nabla_j h,
\eeq
see, for instance, \cite{duzh01}.
\begin{rmk}\label{rmkphham}
The flows \eqref{princflo} of the principal hierarchy are Hamiltonian \cite{duzh01}, as shown by the following considerations. One defines the $1-$forms $\,\theta^{\alpha,n}_i\!\!:=\eta_{ik}X^k_{\alpha,n}$, which are proved to be closed by using \eqref{fdprinchie} together with the flatness of the metric. Therefore, there locally exist functions $H_{\alpha,n}$ such that $\de_iH_{\alpha,n}=\theta^{\alpha,n}_i$, and from \eqref{princ} we have
$$\left(V_{\alpha,n}\right)^i_j=\eta^{ik}\nabla_k\nabla_jH_{\alpha, n+1},$$
which gives the Hamiltonian form of the flows \eqref{princflo}.  Moreover, it has been proved in \cite{duzh01}, that in the semisimple case the functions $H_{\alpha, n}$ satisfy a completeness property, in the sense that they form a complete set in the space of conserved densities of the hierarchy \eqref{princflo}, which are polynomial with respect to one of the flat coordinates.
\end{rmk}
If the Frobenius manifold is semisimple, then in canonical coordinates the flows of the principal hierarchy become diagonal:
$$\frac{\de\, r^i}{\de\, t_{\alpha,n}}=X_{\alpha,n}^i\frac{\de\, r^i}{\de\, x}, \quad \alpha=1,\dots,N, \quad n=0,1,\dots.
$$
In other words, canonical coordinates on a Frobenius manifold are Riemann invariants for the flows of the principal hierarchy, and the components $X^i_{\alpha,n}$ of the vector fields are the characteristic velocities of the corresponding flows. Writing the hierarchy in canonical coordinates, we can then apply the hodograph formula \eqref{hodo} to obtain solutions of the principal hierarchy. Alternatively, one can consider the system
$$x\,\delta^i_j +t\,V^i_j=W^i_j,$$
which is an invariant formulation of the hodograph formula \eqref{hodo}. If $V=V_X$, $W=V_Y$ as in \eqref{htsprod}, for some vector fields $X$, $Y$, then we get
\beq\label{strhod}
c^i_{jk}\left(x\,e^k +t\,X^k-Y^k\right)=0,
\eeq
which shows that the vanishing of the quantities in the bracket defines an implicit solution in any coordinate set. This invariant formulation of the hodograph method in terms of critical points of vector fields has been considered in \cite{lope09}.
\begin{rmk}\label{rmkdegmetr}
We note that although the definition of Frobenius manifold requires the non-degeneracy of the bilinear form $\eta$, the theory of hydrodynamic type systems turns out to be less restrictive. In particular, a local Hamiltonian formalism can be defined also in the degenerate case (see \cite{do93}, Theorem $5.14$), and the principal hierarchy can be built as in Remark \ref{rmkphham}, that is, by using differential forms in place of vector fields.
\end{rmk}
So far, we have considered the direct problem of determining the principal hierarchy of a given Frobenius manifold. The converse problem, of producing a Frobenius manifold starting from a given hydrodynamic type system \eqref{hts} is rather more difficult, and even not always solvable (see \cite{du97} for further detail). We recall here that a necessary condition is the existence for the system \eqref{hts} of a bi-Hamiltonian structure of type \eqref{dnpoisson};  one has then to find a decomposition of \eqref{hts} of the form \eqref{htsprod} (as well as the corresponding unity and Euler vector fields) and to prove the axioms of a Frobenius manifold. If one is able to determine the Frobenius manifold, then the corresponding principal hierarchy form a (complete) set of symmetries of the original system.

\section{dKP equation and Vlasov equations}\label{secvlbe}
\subsection{dKP hierarchy}
The Lax representation of the dispersionless KP hierarchy \cite{tata95} is defined
in terms of the formal power series
\beq\label{genfun}
\la=p+\sum_{k=0}^\infty\frac{A^k}{p^{k+1}},
\eeq
where the variables $A^k$ depend on an infinite set of independent variables $t_n$, with $n>0$, as well as on the spatial variable $x$. The Lax equations are given by
\beq\label{laxnth}
\la_{t_n}=\left\{\Omega_n,\,\la\right\}_{x,p},\qquad \Omega_n=\frac{1}{n}\left(\la^n\right)_{\!+},
\eeq
where the bracket is the canonical Poisson bracket,
\beq\label{canbra}
\left\{h,g\right\}_{x,p}=h_p\,g_x-h_x\,g_p,
\eeq
and $\left(\,\,\,\right)_+$ denotes the polynomial part of the argument. In this setting, every Lax equation can be seen as the generating function of a system of hydrodynamic type with infinitely many dependent variables $A^{k}$, with $k\in\mathbb{N}$.
\begin{exam}
The first non-trivial case, $n=2$, gives
\beq\label{laxbenney}
\la_{t_2}=p\la_x-A^0_x\la_p.
\eeq
This is equivalent to the \emph{Benney moment equation}:
\beq\label{benney}
A^k_{t_2}=A^{k+1}_x+kA^{k-1}A^0_x,\qquad k\in \mathbb{N},
\eeq
named after Benney \cite{be73} who derived it from the study of long nonlinear waves on a shallow perfect fluid with a free surface. For $n=3$ we get
\beq\label{laxrd}
\la_{t_3}=(p^2+A^0)\la_x-(A^0_x\,p+A^1_x)\la_p,
\eeq
to which corresponds the system
\beq\label{momstsym}
A^k_{t_3}=A^{k+2}_x+A^0A^k_x+(k+1)A^kA^0_x+kA^{k-1}A^1_x,\qquad k\in \mathbb{N} .
\eeq
\end{exam}
It is well known that the commutativity of the flows \eqref{laxnth} implies the Zakharov-Shabat (or zero curvature) equations
$$\de_{t_n}\Omega_m-\de_{t_m}\Omega_n=\left\{\Omega_m,\Omega_n\right\}_{x,p},$$
which are systems of PDEs for (2 + 1) independent, and a finite number of dependent variables. In the case $m=2$, $n=3$, and after setting $t_2=y$ and $t_3=t$, one gets the dKP equation \eqref{dkp}.

A Hamiltonian structure for the dKP hierarchy was found by Kupershmidt and Manin \cite{kuma77,kuma78}. Indeed, they wrote the Benney system \eqref{benney} in the form
\beq\label{benneyham1}
A^k_{t_2}=\left\{A^k,\mathcal{H}_2\right\}_1,\qquad \mathcal{H}_2=\int \frac{1}{2}\left(A^2+\left(A^0\right)^2\right)dx,
\eeq
where the Poisson bracket is the \emph{Kupershmidt--Manin bracket}:
\beq\label{kmbracket}
\left\{\mathcal{K},\mathcal{H}\right\}_1=\int\frac{\delta \mathcal{K}}{\delta A^k(x)}\left((k+n)A^{k+n-1}\frac{d}{dx}+nA^{k+n-1}_x\right) \frac{\delta \mathcal{H}}{\delta A^n(x)}dx.
\eeq
We notice that the above bracket is a hydrodynamic type bracket with infinitely many components. In particular, the metric is given by
\beq\label{kmmetric}
\eta^{kn}(A)=(k+n)A^{k+n-1}.
\eeq
All other flows of the dKP hierarchy are also Hamiltonian, with Hamiltonian densities given by the coefficients of the series
$$p=\la-\sum_{k=0}^\infty\frac{H_k}{\la^{k+1}},$$
which is obtained by inverting the series \eqref{genfun} with respect to $p$. The first few of them are
$$H_0=A^0,\quad H_1=A^1,\quad H_2=\frac{1}{2}A^2+\frac{1}{2}\left(A^0\right)^2\quad H_3=\frac{1}{3}A^3+A^0A^1.$$
Furthermore, the Benney equation \eqref{benney} admits a second Hamiltonian formulation,  given by
$$A^k_{t_2}=\left\{A^k,\mathcal{H}_1\right\}_2,\qquad \mathcal{H}_1=\frac{1}{2}\int H_1dx,$$
where the bracket is again of hydrodynamic type with infinitely many components, and it is compatible with the Kupershmidt-Manin bracket. We give here only the metric, which is
\begin{align}\label{metg2}
g^{kn}(A)= \,\,& k\,nA^{k-1}A^{n-1}+(k+n+2)A^{k+n}+\sum_{i=0}^{n-1}(k+i)A^{k+i-1}A^{n-i-1}\nonumber\\
&-\sum_{i=0}^{n-2}(n-i-1)A^{k+i}A^{n-i-2},
\end{align}
further detail can be found in \cite{ku84,gibrai07}. The other flows of the hierarchy are also Hamiltonian with repect to this second Poisson bracket.

\subsection{dKP and Vlasov equations}
The use of the formal series \eqref{genfun}, introduced in the previous section, is to be understood
as an algebraic model for describing the underlying integrable system in a more compact
way. However, to describe the system in more detail we must impose more structure on
$\la$. This has been done --for instance-- in \cite{gits96,gits99}, where $\la$ is defined through the Hilbert transform on the real axis of a suitable function $f$. We now briefly review this approach, without specifying the functional class to which $f$ belongs. Later, in order to construct a suitable Frobenius structure, we will take $f$ in the Schwartz class. We thus define the analogue of the formal series \eqref{genfun} to be the function
\beq\label{lahilbf}
\la(p)=p+\dashint_{-\infty}^{+\infty}\frac{f(q)}{p-q}dq,\qquad p\in\mathbb{R},
\eeq
where $\dashint dp$ denotes the Cauchy principal value integral. The function \eqref{lahilbf} possesses an asymptotic expansion for $p$ at infinity of the form
\beq\label{asymbenney}
\la\sim p+\sum_{k=0}^\infty\frac{A^k}{p^{k+1}},
\eeq
where the coefficients $A^k$, given by 
\beq\label{moments}
A^k=\int_{-\infty}^{+\infty}p^kf(p)dp,\qquad k\in\mathbb{N},
\eeq
are the moments of $f$.  Within this approach, the formal series \eqref{genfun} can be recovered as the asymptotic expansion for $p\mapsto\infty$ of the function \eqref{lahilbf}.  Moreover, due to the definition \eqref{lahilbf} of the function $\la$, to every flow of the dKP hierarchy we can now associate a corresponding equation for the function $f$ \cite{gi81,zak81}. The following examples clarify the situation.  
\begin{exam} The equation 
\beq\label{vlasovbenney}
f_{t_2}=pf_x-A^0_xf_p,
\eeq
leads, under the definition \eqref{lahilbf}, to the dispersionless Lax equation \eqref{laxbenney}. Morever, the moment equations of \eqref{vlasovbenney} are the Benney system \eqref{benney}. 
Analogously, equation
\beq\label{vlasovrd}
f_{t_3}=\left(p^2+A^0\right)f_x-\left(A^0_xp+A^1_x\right)f_p,
\eeq
induces the Lax equation \eqref{laxrd} for $\la$, and the related moment equations turn out to be system \eqref{momstsym}. One can prove that the flows \eqref{vlasovbenney} and \eqref{vlasovrd} commute. Moreover, the dKP equation \eqref{dkp} can be obtained directly from these flows as follows: we set $t_{2}=y$, $t_{3}=t$, and we consider the first few moment equations of \eqref{vlasovbenney} and \eqref{vlasovrd}, that are
$$A^{0}_{y}=A^{1}_{x}, \qquad A^{1}_{y}=A^{2}_{x}+A^{0}A^{0}_{x},\qquad A^{0}_{t}=A^{2}_{x}+2A^{0}A^{0}_{x}.$$
Rearranging, we find the following system of equations
\begin{align*}
A^{1}_{x}&=A^{0}_{y},\\
A^{1}_{y}&=A^{0}_{t}-A^{0}A^{0}_{x},
\end{align*}
which is proved to be compatible, so that $A^{0}$ satisfies the dKP equation \eqref{dkp}.

\end{exam}
\begin{rmk}\label{rmkflaconj}
Comparing equation \eqref{laxbenney} with \eqref{vlasovbenney} and \eqref{laxrd} with \eqref{vlasovrd}, we notice that $f$ and $\la$  are carried along the same characteristics with respect to the flows $t_{2}$ and $t_{3}$. This result can be generalized to all other flows of the dKP hierarchy, and the proof, for every fixed flow, requires elementary manipulations only. However, to the author's best knowledge, a proof of this fact for the complete hierarchy is still missing. 
\end{rmk}
The relation between the $f$ and the $\la$ pictures is not just at the level of the equations, but also at the Hamiltonian level. This is shown by introducing the \emph{Poisson--Vlasov bracket} \cite{mawe82}:
\begin{equation}\label{lpb}
\{\mathcal{G},\mathcal{H}\}_{LP}:=\iint f(x,p) \left\{\frac{\delta\,\mathcal{G}[f]}{\delta f(x,p)},\frac{\delta\, \mathcal{H}[f]}{\delta f(x,p)}\right\}_{x,p}dp\, dx,
\end{equation}
where $\left\{\cdot\,,\,\cdot\right\}_{x,p}$ is the canonical bracket \eqref{canbra}, $\mathcal{G}$, $\mathcal{H}$ are functionals of $f$, and the variational derivatives are defined by the identity
\beq\label{varderd}
\iint \frac{\delta\,\mathcal{H}[f]}{\delta f(x,p)}\Phi(x,p) dp\,dx=\frac{d}{d\epsilon}_{|\epsilon=0}\mathcal{H}[f+\epsilon\Phi],
\eeq
for every suitable test function $\Phi$. 
Hamilton's equations for the bracket \eqref{lpb}:
\begin{equation}\label{genericvlasov}
\frac{\de f}{\de t}=\{f,\mathcal{H}\}_{LP},
\end{equation}
or, equivalently,
\begin{equation}\label{vlasov}
\frac{\de f}{\de t}+\left\{f,\frac{\delta \mathcal{H}}{\delta f}\right\}_{x,p}=0,
\end{equation}
are equations of kinetic type, and more precisely are a class of \emph{Vlasov equations}, which arise in the theories of plasma physics and vortex dynamics. The relation between these equations and the dKP hierarchy relies on the following result \cite{gi81}: if we restrict the bracket \eqref{lpb} to functionals depending on the
moments alone,
$$\mathcal{H}=\mathcal{H}[A^0, A^1,\dots],$$
then the Poisson--Vlasov bracket restricts to the Kupershmidt--Manin bracket \eqref{kmbracket}, namely
$$\left\{\mathcal{G},\mathcal{H}\right\}_{LP}=\left\{\mathcal{G},\mathcal{H}\right\}_1.$$
As a consequence of this fact, every Vlasov equation \eqref{genericvlasov} induces a set of \emph{moment equations}, given by
$$A^k_t=\left\{A^k,\mathcal{H}\right\}_{1}\qquad k=0,1,\dots,$$
where the functions $A^k$ are defined by \eqref{moments}. 
\begin{exam}
The Benney Hamiltonian \eqref{benneyham1}, leads to the Vlasov equation \eqref{vlasovbenney}.
Analogously, the Vlasov equation obtained by the Hamiltonian
$$\mathcal{H}_3=\int\left(\frac{1}{3}A^3+A^0A^1\right)dx,$$
is equation \eqref{vlasovrd}.
\end{exam}

\section{Schwartz functions and tempered distributions}\label{section:sch}
This section is devoted to a brief review of the main properties of Schwartz functions and tempered distributions, in which we can find a sufficiently rich analytical setting for the construction of the Frobenius structure for the dKP equation. Indeed, by taking the function $f$ introduced above in the Schwartz class, we can be more precise on the analytical properties of the objects considered in the previous section. We will be interested on Schwartz functions of one and two variables only.  At the end of the section, we review the well known relation between Schwartz functions of one variable and the Riemann-Hilbert problem on the real axis. 
\subsection{The space $\sch$}
Let $\Phi$ be a function of the real variable $p$. We say that $\Phi$ belongs to the Schwartz class $\mathcal{S}(\mathbb{R})$  if it is $\mathcal{C}^{\infty}-$differentiable and satisfies
$$\underset{p\in\mathbb{R}}{\text{sup}}\,|\,p^k\,\de^{\,m}\Phi(p)\,|<+\infty,$$
for every $k,m\in\mathbb{N}$. The following properties of the space $\sch$ are well known; we refer to the classical references  \cite{sch66,gesh64,resi72} for a more detailed description. The space $\mathcal{S}(\mathbb{R})$ is $k-$linear ($k=\mathbb{R},\mathbb{C}$), and closed with respect to pointwise product and differentiation, meaning that $\phi\,\psi$, $\phi'\!\in\!\mathcal{S}(\mathbb{R})$ if $\phi,\psi\!\in\!\mathcal{S}(\mathbb{R})$. Moreover, we have $\mathcal{S}(\mathbb{R})\subset L^p(\mathbb{R})$, for $1\leq p\leq +\infty$.  The dual space $\mathcal{S}'(\mathbb{R})$ of $\mathcal{S}(\mathbb{R})$ is the space of tempered distributions,  which are defined through the pairing
$$\left<\omega,\Phi\right>=\int_{-\infty}^{+\infty}\!\omega(p)\Phi(p) \,dp,\qquad \omega\in\mathcal{S}'(\mathbb{R}),\, \Phi\in\mathcal{S}(\mathbb{R}).$$
The space $\mathcal{S}'(\mathbb{R})$ can be characterized in the following way: a distribution is tempered if and only if is a finite sum of (weak) derivatives of continuous functions growing at infinity slower than some polynomial. The following examples of tempered distributions will be useful later:
\begin{exam} \label{examproom}Let $\mathcal{O}_M(\mathbb{R})$ denote the set of $\mathcal{C}^\infty(\mathbb{R})$ functions which, together with all their derivatives, grow at infinity slower than some polynomial. We have $\mathcal{S}(\mathbb{R})\!\subset\!\mathcal{O}_M(\mathbb{R})\!\subset\!\mathcal{S}'(\mathbb{R})$, which in particular shows that the Schwartz functions can be seen as tempered distributions. The space $\mathcal{O}_M$ is important for the following reasons: first, given $h\!\in\!\mathcal{O}_M(\mathbb{R})$ and $\Phi\!\in\!\mathcal{S}(\mathbb{R})$ then $h\Phi\!\in\!\mathcal{S}(\mathbb{R})$. This allows one to define the product between $h\!\in\!\mathcal{O}_M$ and $\omega\!\in\!\mathcal{S}'$ by $<\!h\omega,\Phi\!>=<\!\omega,h\Phi\!>$, for every $\Phi\!\in\!\mathcal{S}$. Moreover,  if one defines the convolution of a tempered distribution with a function as 
$$\left(\omega\ast\Phi\right)(p):=\int_{-\infty}^{+\infty}\omega(q)\Phi(p-q)dq,\qquad \omega\!\in\!\mathcal{S}'(\mathbb{R}),\,\Phi\!\in\!\mathcal{S}(\mathbb{R}),$$
then $\omega\ast\Phi\!\in\!\mathcal{O}_M(\mathbb{R})$. The last result is known in the literarure as regularization of a tempered distribution. 
\end{exam}
\begin{exam}\label{deltafundef}
the delta function and its derivatives, which are defined by the conditions
$$\int_{-\infty}^{+\infty}\Phi(q)\delta^{(k)}(p-q)dq=\Phi^{(k)}(p),\qquad k=0,1,2,\dots,$$
for every $\Phi\in\mathcal{S}(\mathbb{R})$, are tempered distributions. They satisfy the identities
\begin{subequations}\label{deltarelation}
\begin{align}
& \delta^{(k)}(p-q)=(-1)^k\delta^{(k)}(q-p),\\
& \Phi(q)\delta^{(k)}(p-q)=\sum_{i=0}^k\binom{k}{i}\Phi^{(i)}(p)\delta^{(k-i)}(p-q).
\end{align}
\end{subequations}
We denote $\,\delta'(p-q)\!=\!\delta^{(1)}(p-q)$, with similar notation for higher derivatives.
\end{exam}
\begin{exam}\label{p.v.}
The singular distribution $\text{p.v.}\tfrac{1}{p}$, defined by
\beq\label{pvintegral}
\left<p.v.\frac{1}{p}\,,\Phi\right>:=\dashint_{-\infty}^{+\infty}\frac{\Phi(p)}{p}dp,
\eeq
see, for instance, \cite{gesh64,pan96}. In order to simplify the notation, the  symbol $p.v.$ will be dropped, that is, we will write $\tfrac{1}{p}$ in place of $p.v.\tfrac{1}{p}$. It will be clear from the context when this quantity has to be understood in the distributional sense.
\end{exam}
We now consider the function $f$ introduced in the previous section, and take $f\in\sch$. This assumption can be weakened, for instance, allowing $f$ to be non-differentiable (or even discontinuous) at some point; however, for simplicity, these generalizations will not be considered in this paper. 

The first consequence of considering a Schwartz function is that all the moments \eqref{moments} are finite, so that the moment equations such as \eqref{benney} make sense. In addition,  the function $\la$ defined in \eqref{lahilbf} belongs to $\mathcal{O}_M(\mathbb{R})$. Indeed, the Cauchy integral appearing in \eqref{lahilbf} can be read as the convolution of $f$ with the distribution of Example \ref{p.v.}. Due to the remark at the end of Example \ref{examproom} we obtain $\la\!\in\!\mathcal{O}_M(\mathbb{R})$. 

\begin{rmk}
The Hilbert transform of a function $\Phi$ is defined as
\beq\label{hilbdef}
\text{Hilb}_p[\Phi]=\frac{1}{\pi}\dashint_{-\infty}^{+\infty}\frac{\Phi(q)}{q-p}dq.
\eeq
Comparing \eqref{hilbdef} with \eqref{lahilbf}, we get
\beq\label{lahilbert}
\la(p)=p-\pi\,\text{Hilb}_p[f].
\eeq
The following classical formulae for the Hilbert transform will be useful later:
\begin{align}
&\text{Hilb}\big[\text{Hilb}\left[\Phi\right]\big]=-\Phi,\\
& \text{Hilb}\big[\Phi_1\text{Hilb}\left[\Phi_2\right]+\text{Hilb}\left[\Phi_1\right]\Phi_2\big]=\text{Hilb}\left[\Phi_1\right]\text{Hilb}\left[\Phi_1\right]-\Phi_1\Phi_2.\label{tricomi}
\end{align}
These conditions hold on the whole real axis provided $\Phi_1$ and $\Phi_2$ belong to $L^p(\mathbb{R})$, $p\!>\!1$, and satisfy the Holder condition
$$||\Phi(p)-\Phi(q)||<C\,|p-q|^\alpha,\qquad 0<\alpha\leq 1,$$
for some constant $C$. In particular, this is true for functions belonging to $\mathcal{S}(\mathbb{R})$.
\end{rmk}
\begin{rmk}
In the definition above, we might allow $f$ to be either real or complex-valued. While the former choice leads to real-valued solutions of the dKP hierarchy, the latter turns out to be important when considering flat coordinates of the Frobenius manifold.  Unless otherwise stated, the results of the present paper hold for both the real and the complex-valued case.
\end{rmk}
We now consider a class of functionals, depending on $f$ but not on its derivatives; we allow these functionals to be nonlocal, and to explicitly depend on $p$.  Given a functional $H[f]$ of this type, its variational derivative $\tfrac{\delta H[f]}{\delta f(p)}$
is defined by the identity
\beq\label{varder}
\left<\frac{\delta H[f]}{\delta f(p)},\Phi(p)\right>=\frac{d}{d\epsilon}_{|\epsilon=0}H[f+\epsilon\,\Phi],
\eeq
for every $\Phi\in\sch$. We restrict ourselves to functionals for which the above identity makes sense, so that the corresponding variational derivative is an element of $\mathcal{S}'(\mathbb{R})$. Note the different notation between the variational derivative in \eqref{varder} and the one in \eqref{varderd}, due to the fact that in the latter case the functionals depend on two variables. Later, and for a special class of functionals, we will describe the relation between these two derivatives.
\begin{exam}\label{examfumom}
Let $H=H(A^0,\dots,A^k)$ be a differentiable function of the moments, and therefore a (nonlocal, in general) functional of $f$. We have
$$\frac{\delta H}{\delta f(p)}=\sum_{i=0}^k\frac{\de H}{\de A^i}\,p^i.$$
Being these functions polynomials, they belong to $\mathcal{O}_M(\mathbb{R})$. 
\end{exam}
\begin{exam}
The tempered distributions of Example \ref{deltafundef} can be written as a variational derivative as
\begin{align*}
\frac{\delta f^{(k)}(p)}{\delta f(q)}&=\delta^{(k)}(p-q),\notag\\
\intertext{while for the distribution \eqref{pvintegral} we have:}
\frac{\delta \la(p)}{\delta f(q)}&=\frac{1}{p-q}.
\end{align*}
Let us explain the last identity. From the definition of variational derivative \eqref{varder}, and recalling \eqref{lahilbert}, we have
$$\left<\frac{\delta \la(p)}{\delta f(q)},\Phi(q)\right>=\frac{d}{d\epsilon}_{|\epsilon=0}\Big(p-\pi\,\text{Hilb}_p[f+\epsilon\,\Phi]\Big)=\dashint_{-\infty}^{+\infty}\frac{\Phi(q)}{p-q}dq,$$
and this is exactly the definition of the singular distribution \eqref{pvintegral}.
\end{exam}

\subsection{The space $\mathcal{S}(\mathbb{R}^{2})$}
From the definition of the Poisson--Vlasov bracket \eqref{lpb} -- and the corresponding Hamilton equations -- it follows that we have to consider the analytical properties of $f$ as a function of both $x$ and $p$.  Although the conditions for the dependence of $f$ through the variable $x$ are less restrictive than the conditions on $p$ (we actually need only $f$ to be smooth on the $(x,p)$ plane and to decay to zero for $|x|$ large), for the sake of simplicity we consider in this paper $f\!\in\!\schd$, the space of smooth and rapidly decreasing functions on the plane. All properties of $\sch$ described above still hold for $\schd$ \cite{sch66}; in particular, we can introduce the dual space $\temd$ through the pairing   
$$\left<\left<\omega,\Phi\right>\right>=\iint \omega(p,x)\Phi(p,x)\,dp\,dx,\qquad \Phi\!\in\!\schd,\,\,\omega\!\in\!\temd,$$
and we can define the space $\omd$ in analogy to the one-variable case.  Furthermore, we introduce a class of functionals, of the two independent variables  $x$ and $p$, as the simplest generalization of the class of functionals in the $p-$variable introduced above. Indeed, let $H[f]$ be one of the above mentioned functionals:  if we let $f$ depend on $x$, then also $H[f]$ becomes a function of $x$,  and  we can define the functional
\beq\label{ourfunc}
\mathcal{H}[f]=\int H[f] dx,
\eeq
provided the integral converges. These functionals are thus local and translational invariant with respect to $x$, while the dependence on $p$ is allowed to be more general. Functionals of the form \eqref{ourfunc} are the analogue, in this setting, of functionals of hydrodynamic type \eqref{funht}. 
\begin{exam}
Let $H(A^0,\dots,A^k)$ be the functional of Example \ref{examfumom}. Then the associated functional 
$$\mathcal{H}[f]=\int H\,\,dx,$$ 
is of the form \eqref{ourfunc}. For a generic choice of the function $H$, the above functional is nonlocal with respect to $p$. Another example is provided by the integral
$$\iint h(f(p,x))\la(p,x)dp\,dx,$$
where $h$ is any differentiable function of $f$ (such that the double integral converges), and where $\la$ is given by \eqref{lahilbf}.  Note that the choice $h=const$ is not allowed, for the corresponding integral diverges.
\end{exam}
Within this construction, the Poisson--Vlasov bracket $\eqref{lpb}$ can now be written as 
$$\left\{\mathcal{G},\mathcal{H}\right\}_{LP}=\left<f,\left\{\frac{\delta\mathcal{G}}{\delta f},\frac{\delta\mathcal{H}}{\delta f}\right\}_{x,p}\right>,$$
where the variational derivative is defined by \eqref{varderd}. 
Notice the difference between \eqref{varder} and \eqref{varderd},  due to the fact that the former is computed at $x$ fixed. It is not difficult to show that, for functionals of the form \eqref{ourfunc}, the two variational derivatives are related by
\beq\label{pvbra}
\frac{\delta\,\mathcal{H}}{\delta f(p,x)}=\frac{\delta H}{\delta f(p)},
\eeq
which is the analogue of \eqref{fvdientity}. 
The Poisson--Vlasov bracket \eqref{lpb} of the functionals $\mathcal{G}$ and $\mathcal{H}$ is well defined provided the canonical Poisson bracket of the corresponding variational derivatives belongs to $\mathcal{S}'(\mathbb{R}^2)$. This requirement is satisfied, for instance,  if $\tfrac{\delta\mathcal{G}}{\delta f}, \tfrac{\delta\mathcal{H}}{\delta f}\in\mathcal{O}_M(\mathbb{R}^2)$, and this is equivalent to say that the canonical Poisson bracket gives to $\omd$ a Lie algebra structure, with \eqref{pvbra} playing the role of the associated Lie-Poisson bracket. However, the formula still makes sense if one of the arguments, say $\tfrac{\delta\mathcal{G}}{\delta f}$, belongs to $\mathcal{S}'(\mathbb{R}^{2})\setminus\omd$. The important choice $\mathcal{G}\!=\!f(q,\tau)$, which gives Hamilton's equations \eqref{genericvlasov}, \eqref{vlasov}, has variational derivative
$$\frac{\delta f(q,\tau)}{\delta f(p,x)}=\delta(q-p)\,\delta(\tau-x),$$
which is an element of $\temd$ but does not belong to $\omd$.

\subsection{Riemann--Hilbert problem}
The definition of the function \eqref{lahilbf} in place of the formal series \eqref{genfun}, it is not only useful from an analytical viewpoint, but leads also to a nice geometrical construction, relating the above quantities to the solution of a scalar Riemann-Hilbert problem on the real axis. Here all functions are considered at $x$ fixed. Since $f\in\mathcal{S}(\mathbb{R})$ \footnote{the same result holds for weaker conditions, see for instance \cite{mus53,tit86}.}, then the functions
$$\Phi_\pm(p)=\mp f(p)+i\,\text{Hilb}_p[f],\qquad p\in\mathbb{R},$$
are the boundary values on the real axis of a complex function $\Phi(p)$, which is analytic for $p\in\mathbb{C}\setminus\mathbb{R}$. In our case, we thus have
$$\Phi_\pm(p)=\mp f(p)-\frac{i}{\pi}\left(\la(p)-p\right),\qquad p\in\mathbb{R},$$
which, with $\la$ given by \eqref{lahilbf}, is the unique solution of the scalar Riemann--Hilbert problem
\begin{align*}
&\Phi_+(p)-\Phi_-(p)=-2 f(p),\quad p\in\mathbb{R}\\
&\Phi_\pm(p)\mapsto 0, \qquad p\mapsto \infty.
\end{align*}
By slightly modifying the above problem, we now define the pair of functions
\beq\label{rhpair}
\tilde{\Phi}_\pm(p)=\mp\pi f(p)-i\,\la(p),\qquad p\in\mathbb{R},
\eeq
which are solutions of a Riemann--Hilbert problem similar to the former, but with a different normalization at infinity. The advantage of this definition is that the funtions \eqref{rhpair} --being linear combinations of $f$ and $\la$-- satisfy equations \eqref{laxbenney} and \eqref{laxrd} (compare with Remark \ref{rmkflaconj}).
In addition, $\tilde{\Phi}_\pm$ have the same asymptotic expansion \eqref{asymbenney}, for large $p$, as $\la$. The function $\tilde{\Phi}_+$ or , more precisely, the associated plane curve $\gamma:\mathbb{R}\rightarrow\mathbb{R}^2,$ given by
\beq\label{planecurve}
\gamma(p)=\left(-\pi f(p),-\la(p)\right),\qquad p\in\mathbb{R},
\eeq
will be useful in Section \ref{secspeccoo}, when considering canonical coordinates for the Frobenius manifold of dKP.

\begin{rmk}
By considering yet another pair of functions: $\psi_\pm:=i\,\tilde{\Phi}_\pm$, we get
$$\psi_\pm(p)=\la(p)\mp i\pi f(p).$$
This is the classical decomposition, considered in relation with the dKP hierarchy, for example, in the papers \cite{giko94,gits99,giyu00}. In the case of reductions of the dKP hierarchy, the analytic continuation of $\psi_+$ in the upper half plane is conformal map, solution of a system of chordal Loewner equations.
 \end{rmk}

Due to the above considerations, for the construction of the Frobenius manifold it is possible to use -- instead of the function $f$  -- either  $\la$ or one of the functions $\tilde{\Phi}_{\pm}$.  Although this approach is closer to the one considered in \cite{cadume09}, the continuous index approach provided in this paper for the dKP equation is better suited for the $f-$picture.

\section{A Frobenius manifold for dKP}\label{sec:from}
\subsection{Vlasov equation as hydrodynamic type system}
We begin now the construction of a Frobenius manifold, which we want to relate to a class of Vlasov equations \eqref{vlasov}, and in particular with equations \eqref{vlasovbenney} and \eqref{vlasovrd}, which give dKP. Our approach is to consider these equations as a continuous--indexed hydrodynamic type system, and to proceed with the construction of the Frobenius manifold in full analogy with the finite dimensional case. This procedure has already been considered in \cite{gibrai07} where the Haantjes tensor (or, better, its continuous indexed analogue) has been computed for equation \eqref{vlasovbenney}, as well as for more general Vlasov equations. The idea is to consider the variable $p$, appearing in the function $f(p,x)$, as a continuous parameter, rather than as an independent variable. In other words, we take the set 
$$\left\{f(p),\,p\in\mathbb{R}\right\},$$
to formally play the role of the coordinates $u^1, \dots, u^N$ in the finite dimensional case. Hydrodynamic type systems take thus the form
\beq\label{conhts}
f_t(p)=\int V\tbinom{p}{q}f_x(q)dq,
\eeq
where the kernel  $V\tbinom{p}{q}$ is a functional of $f$. In order to remain closer to the notation of Section \ref{secfrob}, the dependence on $x$ of the function $f$ and its derivatives will be omitted.
More precisely, we denote 
$$f(p):=f(x,p),\quad f'(p):=\frac{\de f(x,p)}{\de p},\quad f_x(p):=\frac{\de f(x,p)}{\de x},$$
with similar notation for higher derivatives as well as for other functions. We remark that, for every fixed $x$, the above functions all belong to $\sch$. Therefore, we can define equations of type \eqref{conhts} by looking at a class of linear operators
\beq\label{vss}
V\tbinom{p}{q}:\mathcal{S}(\mathbb{R})\longrightarrow\mathcal{S}(\mathbb{R}),
\eeq
thus identifying the vector space $\sch$ with the tangent space. The dual space $\tem$ plays therefore the role of cotangent space; we anticipate here that in order to give a precise meaning to the objects involved in the construction of the Frobenius structure,  we will need to consider suitable vector subspaces of both $\sch$ and $\tem$.
\begin{exam}
The simplest choice
\beq
V\tbinom{p}{q}=\delta(p-q),
\eeq
give rise to the system $f_{t}(p)=f_{x}(p)$, which is  the first element of the dKP hierarchy. A more interesting example is given by choosing in \eqref{conhts}
\beq\label{vlasovbenneymat}
V\tbinom{p}{q}=p\,\delta(p-q)-f'(p),
\eeq
from which we get the Vlasov--Benney equation \eqref{vlasovbenney}. Another example is given by the kernel
\beq\label{vlasovrdmat}
V\tbinom{p}{q}=\left(p^2+A^0\right)\delta(p-q)-(p+q)f'(p),
\eeq
which gives equation \eqref{vlasovrd}. We note that all these examples are of type \eqref{vss}.
\end{exam}

\subsection{Compatible flat metrics}
We have seen in Section \ref{secvlbe} that the Benney system has a bi-Hamiltonian structure with respect to two Poisson brackets of hydrodynamic type. Therefore,  the two natural candidates to become the first metric and the intersection form of the Frobenius manifold are, respectively, the flat metrics \eqref{kmmetric} and \eqref{metg2}. We need to write these metrics in the continous formalism. However, it is a result of \cite{gibrai07} that the Poisson--Vlasov bracket \eqref{lpb}, which is the form of the Kuperschmidt--Manin bracket \eqref{kmbracket} written in the $f-$picture, can be written as bracket of hydrodynamic type with continuous indices, with the contravariant metric given by
\beq\label{stmetric}
\eta^{(p\,q)}=-f'(p)\delta(p-q).
\eeq
This is the continuous form, in the $f-$picture, of the metric \eqref{kmmetric}. Indeed, \eqref{stmetric} defines a bilinear map on the vector space $\om\subset\tem$ given by
\beq\label{stcontracomp}
\eta(h^1,h^2)=\iint h^1_{(p)}\eta^{(p,q)}h^2_{(q)}dp\,dq=-\left< h^1,f'h^2\right>,
\eeq
for every $h^{1},\,h^{2}\in\om$. This bilinear form is strictly related with the linear operator
$$\rho:\mathcal{O}_M(\mathbb{R})\longrightarrow\mathcal{S}(\mathbb{R}),\qquad \rho(h)(p)=f'(p)h(p),$$
so that we can write
$$\eta(h^1,h^2)=\langle h^1,\rho(h^2)\rangle.$$
\begin{rmk}
The map $\rho$ is never onto and, in general, not $1\!-\!1$ either. It is not difficult to show that
$$\ker{\rho}=\left\{h\in\mathcal{O}_M(\mathbb{R}):\mu\big(\text{supp}(h)\cap\text{supp}(f')\big)=0\right\},$$
where $\mu$ denotes the Lebesgue measure on $\mathbb{R}$. Since the elements of $\om$ are continuous functions, we have that the $\ker{\rho}$ is non-trivial if and only if $f'\!=\!0$ on a set of positive measure.  
\end{rmk}
We denote $\mathcal{V}=\im{\rho}$. This is a vector subspace of $\sch$, whose elements are of the form $X^{(p)}=h(p)f'(p)$, for some $h\in\om$. On the space $\mathcal{V}$, we can define the inverse metric of $\eta$, with components
\beq\label{fstmet}
\eta_{(p\,q)}=-\frac{1}{f'(p)}\delta(p-q),
\eeq
which acts on vectors $X_{1},\,X_{2}\in\mathcal{V}$ as follows:
$$\eta(X_{1},X_{2})=\iint X_{1}^{(p)}\eta_{(p,q)}X_2^{(q)}dp\,dq=-\int\frac{X_1^{(p)}X_2^{(p)}}{f'(p)}dp.$$
By construction, the above bilinear form is well defined on $\mathcal{V}$.  Following \cite{gibrai07}, we now consider an analogue of the
finite dimensional, differential geometric objects, which is obtained by replacing partial derivatives with variational one, and sums over repeated indices with integrals. Remarkably, this construction turns out to be consistent with the rest of the theory. For instance, we define the Christoffel symbols of the metric \eqref{fstmet} by taking the usual finite dimensional formula, and we obtain  \cite{gibrai07}:
\beq\label{christ}
\Gamma\tbinom{p}{q\,r}=\frac{1}{f'(q)}\delta'(q-p)\delta(q-r).
\eeq
Since the metric \eqref{fstmet} is not constant with respect to $f$, we thus say that the $f(p)$ are not flat coordinates for the metric. Furthermore, we denote by $\nabla$ the corresponding covariant derivative, which we require to act on elements of $\mathcal{V}$ as
$$\nabla_{(q)}X^{(p)}=\frac{\delta X^{(p)}}{\delta f(q)}+\int \Gamma\tbinom{p}{q\,r}X^{(r)}dr=\frac{\delta X^{(p)}}{\delta f(q)}+\delta'(q-p)X^{(p)},$$
and on $h\in\om$ as
$$\nabla_{(p)}h_{(q)}=\frac{\delta h_{(q)}}{\delta f(p)}-\int \Gamma\tbinom{r}{p\,q}h_{(r)}dr=\frac{\delta h_{(q)}}{\delta f(p)}-\delta(p-q)\frac{\de h_{(p)}}{\de p}.$$
The two actions can be proved to be consistent by putting $X\!=\!hf'$ and using the properties \eqref{deltarelation} of delta functions. The action of $\nabla$ on more general tensor fields is defined in analogy with the finite dimensional case; for instance, the well known formula
$$\nabla_{(p)}\eta_{(q,r)}=\frac{\delta \eta_{(q,r)}}{\delta f(p)}-\int \Gamma\tbinom{s}{p\,q}\eta_{(s,r)}dr-\int \Gamma\tbinom{s}{p\,r}\eta_{(q,s)}dr=0,$$
holds in this continuous--index setting. The second metric \eqref{metg2} can also be written in this formalism as
\beq\label{ndmetric}
g^{(p\,q)}=f'(p)f'(q)-\frac{f(p)f'(q)-f(q)f'(p)}{p-q}+\delta(p-q)\Big(f(p)\la'(p)-f'(p)\la(p)\Big),
\eeq
and the corresponding Christoffel symbols can be found in \cite{gibrai07}, where the metrics \eqref{stmetric} and \eqref{ndmetric} are also proved to have zero Riemann curvature tensor and to be compatible. In particular, we will take \eqref{stmetric} to be the first metric of the Frobenius manifold; in order to prove that \eqref{ndmetric} is the intersection form, we will have to introduce suitable structure constants of the algebra and the Euler vector field.

\subsection{Structure constants of the algebra}
The next step in the construction of the Frobenius manifold is to consider a product of vectors of the form
\begin{gather}\label{prodconf}
\left(X\circ Y\right)^{(p)}=\iint X^{(q)}c\tbinom{p}{qs}Y^{(s)}dqds,
\end{gather}
where $X$, $Y$ belong to $\mathcal{V}$. We consider the quantities
\beq\label{fconst}
c\binom{p}{q\,r}=\frac{\delta(p-r)}{p-q}-\frac{f'(p)}{f'(q)}\frac{\delta(q-r)}{p-q}+\frac{\delta(p-q)}{q-r}-\frac{\la'(p)}{f'(p)}\delta(q-r)\delta(q-p),
\eeq
where $\la(p)$ is given by \eqref{lahilbf} and, consequently,
$$\la'(p)=1+\dashint_{-\infty}^{+\infty}\frac{f'(q)}{p-q}dq.$$
It is not difficult to show that $\mathcal{V}$ is closed under the product \eqref{prodconf}, \eqref{fconst}. Moreover, we have the following
\begin{thm} The quantities \eqref{fconst} give to $\mathcal{V}$ the structure of a symmetric, associative algebra, which is compatible with the metric \eqref{stmetric} and with its Levi-Civita connection. Namely, \eqref{fconst} satisfy the following conditions:
\begin{itemize}
 \item Symmetry
$$c\dbinom{p}{q\,r}=c\dbinom{p}{r\,q}.$$
\item Associativity
$$\int c\dbinom{p}{q\,s}c\dbinom{s}{l\,r}ds=\int c\dbinom{p}{l\,s}c\dbinom{s}{q\,r}ds.$$
\item Compatibility with the metric
$$\int \eta_{(p\,q)}c\dbinom{q}{r\,s}dq=\int \eta_{(r\,q)}c\dbinom{q}{p\,s}dq.$$
\item Compatibility with the connection
$$\nabla_{(l)}c\dbinom{p}{q\,r}=\nabla_{(q)}c\dbinom{p}{l\,r}.$$
\end{itemize}
\end{thm}
\pf The first three conditions are proved by using the identities \eqref{deltarelation} for the Dirac delta function and its derivatives. For the last condition, after computing the quantity
$$\nabla_{\!(l)}c\binom{p}{q\,r}=\frac{\delta c\tbinom{p}{q\,r}}{\delta f(l)}+\!\!\int\!\Gamma\tbinom{p}{l\,s} c\tbinom{s}{q\,r}ds-\!\!\int\!\Gamma\tbinom{s}{l\,q}c\tbinom{p}{s\,r}ds-\!\!\int\!\Gamma\tbinom{s}{l\,r}c\tbinom{p}{q\,s}ds,$$
one can prove its symmetry with respect to the indices $l$ and $q$ by using similar methods as in the first three conditions.
\epf
As a consequence of the above theorem, the metric \eqref{stmetric} and the structure constants \eqref{fconst} satisfy the axioms appearing in the definition of a Frobenius manifold. In order to complete the construction, we still have to determine the unity of the algebra and the Euler vector field.
\begin{prop}
The vector field $e\in\mathcal{V}$ with coefficients
\beq\label{evf}
e^{(p)}=-f'(p),
\eeq
is the unity of the algebra \eqref{fconst}, and it is flat with respect to the connection \eqref{christ}.
\end{prop}
\pf
By applying $e$ to the product \eqref{fconst}, we get
\begin{align*}
&\int c\tbinom{p}{q\, r}e^{(r)}dr=-\frac{f'(p)}{p-q}+\frac{f'(p)}{p-q}-\delta(p-q)\left(\int\frac{f'(r)}{p-r}dr-\la'(p)\right)=\delta(p-q).
\end{align*}
which shows that $e$ is the identity of the algebra. Moreover, we have
$$\nabla_{(q)}e^{(p)}=\frac{\delta\, e^{(p)}}{\delta f(q)}+\int\!\Gamma\tbinom{p}{q\,s}e^{(s)}ds=-\delta'(p-q)-\delta'(q-p)=0,$$
and therefore $e$ is a flat vector field with respect to the connection $\nabla$.
\epf
The bilinear operator \eqref{prodconf} with the structure constants \eqref{fconst} acts on pair of vectors in the space $\mathcal{V}$, which is a vector subspace of $\sch$. Hence, for every $X\in\mathcal{V}$ we can define the linear map
$$V_{X}:\mathcal{V}\longrightarrow\mathcal{V},$$
by the formula
$$V_{X}\tbinom{p}{q}=\int c\tbinom{p}{q\,r}X^{(r)}dr.$$
Since $X\in\mathcal{V}$, it is of the form $X^{(p)}=h_{(p)}f'(p)$, for some $h\in\om$. Hence, we have
$$V_{X}\tbinom{p}{q}=\frac{h_{(p)}-h_{(q)}}{p-q}f'(p)+\delta(p-q)\left(\int\frac{h_{(r)}f'(r)}{p-r}dr-h_{(p)}\la'(p)\right),$$
from which it follows that the operator $V_{X}$ makes sense when applied to any vector in $\sch$. We can thus extend $V_{X}$ to a map of type \eqref{vss}, and in particular, we can apply it to the vector $f_{x}(p)\in\sch\setminus\mathcal{V}$. The evolutionary equation
\beq\label{evocon}
f_{t}(p)=\int V_{X}\tbinom{p}{q}f_{x}(q)dq,
\eeq
is thus well defined, and $f_{t}$ belongs to $\sch$. The following example shows that the product \eqref{fconst} is associated with the flows of the dKP hierarchy. 
\begin{exam}
Consider the vector field
\beq\label{xvf}
X^{(p)}=-pf'(p)\in\mathcal{V}.
\eeq
Then, we have:
\begin{align*}
V_X\tbinom{p}{q}:=&\int c\tbinom{p}{q\, r}X^{(r)}dr\\
=&\,\,-\frac{pf'(p)}{p-q}+\frac{qf'(p)}{p-q}-\delta(p-q)\left(\int\frac{rf'(r)}{p-r}dr-p\la'(p)\right)\\
=& \,-f'(p)-\delta(p-q)\left(\int\frac{(r-p)f'(r)}{p-r}dr-p\right)\\
=&\,p\,\,\delta(p-q)-f'(p),
\end{align*}
which is exactly \eqref{vlasovbenneymat}. By a similar calculation, we can prove that the vector field
\beq\label{yvf}
Y^{(p)}=-\left(p^2+2A^0\right)f'(p)\in\mathcal{V},
\eeq
applied to \eqref{fconst} gives \eqref{vlasovrdmat}.
\end{exam}

\subsection{Euler vector field}
We now consider the Euler vector field associated with the Frobenius manifold. Since we already have two compatible flat metrics, namely \eqref{stmetric} and \eqref{ndmetric},  we look for a vector field $E$ satisfying the condition
\beq\label{contintform}
g^{(p\,q)}=\iint E^{(s)}\eta^{(p\,r)}c\tbinom{q}{rs}dsdr,
\eeq
thus assuming that the metric $g$ is the intersection form. Expanding the right hand side of the above formula we find
$$g^{(p,q)}=\frac{f'(p)E^{(q)}-f'(q)E^{(p)}}{p-q}+\delta(p-q)\left(\la'(p)E^{(p)}-f'(p)\int\frac{E^{(r)}}{p-r}dr\right),$$
and, comparing with \eqref{ndmetric}, we find that the vector field
\beq\label{euler}
E^{(p)}=f(p)-pf'(p),
\eeq
is a solution of \eqref{contintform}. Moreover, we have:
\begin{thm}
The vector field \eqref{euler} satisfies the following conditions,
\begin{align*}
&\nabla_p\nabla_q E^{(r)}=0,\\
&\!\int\!\!E^{(s)}\,\frac{\delta c\tbinom{p}{qr}}{\delta f(s)}ds-\!\int\!\!c\tbinom{s}{qr}\,\frac{\delta E^{(p)}}{\delta f(s)}ds+\!\!\int\!\frac{\delta E^{(s)}}{\delta f(q)}\,c\tbinom{p}{sr} ds+\!\!\int\!\frac{\delta E^{(s)}}{\delta f(r)}\,c\tbinom{p}{qs}ds=c\tbinom{p}{qr},\\
&\int E^{(r)}\frac{\delta\eta_{(p\,q)}}{\delta f(r)}dr+\int\eta_{(r\,q)}\frac{\delta E^{(r)}}{\delta f(p)}dr+\int\eta_{(p\,q)}\frac{\delta E^{(r)}}{\delta f(q)} dr=3\,\eta_{(p\,q)},
\end{align*}
and it is therefore the Euler vector field of the Frobenius manifold. Moreover, the Frobenius manifold has charge $d=-1$.
\end{thm}
\pf
The above conditions are the continuous index versions of the conditions for the Euler vector field, written in components. They can all be verified by a direct calculation, we prove here only the last. Since we have
$$\frac{\delta E^{(p)}}{\delta f(q)}=\delta(p-q)-p\delta'(p-q),\quad \frac{\delta\eta_{(p\,q)}}{\delta f(r)}=\frac{1}{f'(p)^2}\delta'(p-r)\delta(p-q),$$
substituting in the right hand side of the above formula we get
\begin{align*}
&\!\int\!\!\left(f(r)-rf'(r)\right)\frac{\delta'(p-r)\delta(p-q)}{f'(p)^2}dr-\int\!\! \frac{1}{f'(r)}\delta(r-q)\left(\delta(r-p)-r\delta'(r-p)\right) dr\\
&-\!\int \frac{1}{f'(p)}\delta(p-r)\left(\delta(r-q)-r\delta'(r-q)\right) dr\\
=&-p\,\frac{f''(p)}{f'(p)^2}\delta(p-q)-\frac{2}{f'(p)}\delta(p-q)+\left(\frac{p}{f'(p)}-\frac{q}{f'(q)}\right)\delta'(p-q)\\
=&-\frac{3}{f'(p)}\,\delta(p-q)=3\,\eta_{(p\,q)},
\end{align*}
where we used the properties \eqref{deltarelation} of the delta function. Moreover, this calculation fixes the charge $d$ of the Frobenius manifold to be equal to $-1$.
\epf
\begin{cor}
The metric $g$ given by \eqref{ndmetric} is the intersection form of the Frobenius manifold.
\end{cor}
\begin{rmk}
For a generic choice of $f$, we note that $E\in\sch\setminus\mathcal{V}$, and therefore we cannot define a flow of the form \eqref{evocon} using the Euler vector field. This fact has a partial counterpart in finite dimension, where the flow generated by the Euler vector field via \eqref{htsprod} is not a member of a hydrodynamic type hierarchy, unless the whole hierarchy is degenerate.
\end{rmk}

\subsection{Potential of the Frobenius manifold}
Since the coordinates $f(p)$ are not flat coordinates for the metric \eqref{fstmet}, in order to find the potential of the Frobenius manifold we use the invariant formulation \eqref{potanycoo}, written in the continuous index form. We introduce the quantities
$$c_{(p\,q\,r)}:=\int\eta_{(p\,s)}c\tbinom{s}{q\,r}ds=-\frac{1}{f'(p)}\,c\tbinom{p}{q\,r},$$
and we have the following
\begin{thm} The functional
\beq\label{potential}
F=\frac{1}{2}\iint \log{|p-q|}f(p)f(q)dp\,dq+\frac{1}{2}\int p^2f(p)dp.
\eeq
satisfies the condition
$$\nabla_{(p)}\nabla_{(q)}\nabla_{(r)}F=c_{(p\,q\,r)},$$
and is therefore the potential of the Frobenius manifold.
\end{thm}
\pf
We verify the theorem by computing the third covariant derivative of the functional $F$. First, we get
$$\nabla_{(r)}F=\frac{\delta F}{\delta f(r)}=\int\log{|r-\alpha|}f(\alpha)d\alpha+\frac{1}{2}\,r^2.$$
Then,
\begin{align*}
\nabla_{(q)}\nabla_{(r)}F&=\frac{\delta}{\delta f(q)}\,\nabla_{(r)}F-\int\Gamma\tbinom{s}{qr}\nabla_{(r)}Fds\\
&=\log{|r-q|}-\frac{1}{f'(q)}\delta(q-r)\frac{\de}{\de q}\left(\int\log{|q-\alpha|}f(\alpha)d\alpha+\frac{1}{2}q^2\right)\\
&=\log{|r-q|}-\frac{\la(q)}{f'(q)}\delta(q-r),
\end{align*}
and, denoting $\theta_{(q\, r)}:=\nabla_{(q)}\nabla_{(r)}F$, we obtain
\begin{align*}
\nabla_{(p)}\theta_{(q\, r)}=&\frac{\delta\,\theta_{(q\,r)}}{\delta f(p)}-\int\Gamma\tbinom{s}{pq}\theta_{(s\,r)}ds-\int\Gamma\tbinom{s}{pr}\theta_{(q\,s)}ds\\
=&-\frac{1}{f'(q)}\frac{\delta(q-r)}{q-p}+\frac{\la(q)}{f'(q)^2}\delta(q-r)\delta'(q-p)\\
&-\frac{1}{f'(p)}\delta(q-p)\frac{\de\theta_{(p\,r)}}{\de p}-\frac{1}{f'(p)}\delta(p-r)\frac{\de\theta_{(q\,p)}}{\de p}\\
=&\frac{1}{f'(q)}\frac{\delta(q-r)}{p-q}+\frac{\la(q)}{f'(q)^2}\delta(q-r)\delta'(q-p)\\
&-\frac{1}{f'(p)}\frac{\delta(p-q)}{p-r}+\frac{\la'(p)}{f'(p)^2}\delta(q-p)\delta(p-r)\\
&-\frac{\la(p)}{f'(p)^3}\delta(q-p)\delta(p-r)+\frac{\la(p)}{f'(p)^2}\delta(q-p)\delta'(p-r)\\
&-\frac{1}{f'(p)}\frac{\delta(p-r)}{p-q}-\frac{\la(q)}{f'(p)f'(q)}\delta(p-r)\delta'(q-p).
\end{align*}
By using identities of delta functions, we finally get to
\begin{align*}
\nabla_{(p)}\nabla_{(q)}\nabla_{(r)}F=&\frac{1}{f'(q)}\frac{\delta(q-r)}{p-q}-\frac{1}{f'(p)}\frac{\delta(p-r)}{p-q}\\
&-\frac{1}{f'(p)}\frac{\delta(p-q)}{p-r}+\frac{\la'(p)}{f'(p)^2}\delta(q-p)\delta(p-r),
\end{align*}
which is exactly $c_{(p\,q\,r)}$. This completes the proof.
\epf
\noindent
If we take $f\leq 0$, that is if
$$d\mu(p)=-f(p)dp,$$
is a measure, absolutely continuous with respect to the Lebesgue measure, then the potential $F$ can be written as
\beq\label{partition}
F=-\frac{1}{2}\left(\iint \log{|p-q|^{-1}}d\mu(p)d\mu(q)+\int p^2d\mu(p)\right).
\eeq
The above function is an example of a  logarithmic potential with external field \cite{sato97}.  In particular, the quantity inside the bracket appears in random matrix theory \cite{de99}, when considering the equilibrium measure for the large $N$ limit of the partition function for the Gaussian Unitary Ensemble. The choice of the multiplication constant $-\frac{1}{2}$ is put here only for convenience in the computation of the structure constants. We remark that we do not require $d\mu$ to be a probability measure, and that the dependence on the times of the hierarchy is implicitly contained in the measure $d\mu$. As a consequence of the above theorem, we also note that the identity
\beq\label{potlaide}
\frac{d}{dp}\frac{\delta F}{\delta f(p)}=\la(p),
\eeq
holds.

\begin{rmk}
The relation between dispersionless integrable systems and random matrices has already been considered, for instance, in Wiegmann, Zabrodin \cite{wiza00} and Elbau and Felder \cite{elfe05}, in relation with the dispersionless limit of the $2D$ Toda hierarchy. This suggests the possibility of extending potential \eqref{potential} to a wider class of external fields, and to consider the corresponding Frobenius structure. Preliminary calculations show that the structure constants obtained in this way are not compatible with the metric \eqref{fstmet}, so that the extension is not straightforward, but it requires a more detailed study.
\end{rmk}
\begin{rmk}
It would be interesting to compare the potential \eqref{potential} with the one found by Zabrodin in \cite{zab09}, where he considers the dKP hierarchy in relation with the growth of \lq fat slits\rq on the upper complex half plane (see also \cite{bmrwz01}). In principle, and except for some technical difference, it should be possible to understand this relation in terms of the Riemann--Hilbert problem associated to dKP. However, this approach seems not so easy to apply at this stage;  further work is needed in this direction.
\end{rmk}

\section{Principal Hierarchy}\label{secprinchie}
\subsection{Flat vector fields}
As explained in Section \ref{secfrob}, to every Frobenius manifold one can associate a set of commuting flows of hydrodynamic type, known as the principal hierarchy. This is done by considering the flat vector fields for the first metric, and then applying the recursive procedure \eqref{princ}. In our case, the first step is thus  to find flat vector fields for the metric \eqref{stmetric}. In analogy with the finite dimensional case, one is expected to find in this case inifinitely many flat vector fields,  parametrized by a continuous real index; however, we consider here a larger class, parametrized by an arbitrary function of one variable. Indeed, let us introduce hydrodynamic type functionals of the form
\beq\label{casimirdens}
H_{h,0}=\int h(f)\,dp,
\eeq
where $h$ is a function of one variable, and the integral
\beq\label{casimir}
\mathcal{H}_{h,0}=\int H_{h,0}\,dx,
\eeq
is supposed to converge. For instance, we might take $h$ so that $h(f)\in\schd$. It is well known that the functionals \eqref{casimir} are Casimirs of the Lie--Poisson bracket \eqref{lpb}; by using the densities $H_{h,0}$ we can introduce the vector fields
\beq\label{flatvfdkp}
X^{(p)}_{h,0}=\int \frac{\delta H_{h,0}}{\delta f(q)}\,\eta^{(q,p)}dq=-h'(f(p)) f'(p),
\eeq
which belong to $\mathcal{V}$, and prove the following
\begin{prop}
The vector fields \eqref{flatvfdkp} are flat vector fields for the Levi-Civita connection $\nabla$ of the metric \eqref{stmetric}, that is, they satisfy the condition
$$\nabla_{(q)}X^{(p)}_{h,0}=0.$$
\end{prop}
\pf
We have
\begin{align*}
\nabla_{(q)}X^{(p)}_{h,0}=&\frac{\delta X^{(p)}_{h,0}}{\delta f(q)}+\int\Gamma\tbinom{p}{qr}X^{(r)}_{h,0}dr\\
=&-h'(f(p))f'(p)\delta(p-q)-h(f(p))\delta'(p-q)-\delta'(q-p)h(f(q))\\
=&\,0,
\end{align*}
the last identity due to the properties of the delta function derivatives.
\epf
\noindent
We note that the flat vector field obtained by choosing $h(f)=f$ in \eqref{flatvfdkp} is the unity $e^{(p)}$ of the algebra \eqref{fconst}. We now introduce the primary flows of the principal hierarchy as the PDEs
\beq\label{primflodkp}
\de_{t_{h,0}}f(p)=\int V_{h,0}\tbinom{p}{q}f_x(q)dq,\qquad V_{h,0}\tbinom{p}{q}:=\int c\tbinom{p}{qr}X^{(r)}_{h,0}.
\eeq
We remark that in the classical construction of a principal hierarchy, the use of the term \lq primary flow\rq is more stringent that the one we consider here, for it is related to a special choice of the flat vector fields. Nevertheless, since the flows \eqref{primflodkp} generate by recursion the rest of the hierarchy, it is convenient to name these PDEs primary flows.
\begin{prop}
The flows \eqref{primflodkp} are Hamiltonian of the form
\beq\label{hamprimflo}
\de_{t_{h,0}}f(p)=\left\{f(p),\int H_{h,1}\,dx\right\}_{LP},
\eeq
where the Poisson bracket is the Poisson--Vlasov bracket \eqref{lpb}, and the Hamiltonian is given by
\beq\label{hh1}
H_{h,1}=\int h(f(p))\la(p)\,dp.
\eeq
\end{prop}
\pf
It is well known \cite{duzh01} that the primary flows of the principal hierarchy are Hamiltonian with respect to the first Poisson bracket, and with  Hamiltonian density given --in our formalism-- by the formula
$$H_{h,1}=\int X^{(p)}_{h,0}\frac{\delta F}{\delta f(p)}dp,$$
where the functional $F$ si the potential \eqref{potential} of the Frobenius manifold. Therefore, we have
$$H_{h,1}=-\int h'(f(p)) f'(p)\frac{\delta F}{\delta f(p)}=\int h(f(p))\frac{d}{dp}\frac{\delta F}{\delta f(p)}dp=\int h(f(p))\la(p)\,dp,$$
where in the last identity we have used \eqref{potlaide}. Moreover, since
$$\frac{\delta H_{h,1}}{\delta f(p)}=h'(f(p))\la(p)-\int\frac{h(f(q))}{p-q}dq,$$
one can directly  prove that equations \eqref{primflodkp} and \eqref{hamprimflo} coincide.
\epf
\subsection{Recursive relation} 
Let us now consider the recursive relation \eqref{princ}. Looking at the construction of the primary flows, and in particular at the functionals \eqref{casimirdens} and \eqref{hh1}, it seems reasonable to consider functionals of the form
\beq\label{princhamden}
H_{h,n}=\frac{1}{n!}\int h(f(p))\la(p)^n\,dp,
\eeq
and to use these functionals in order to construct the vector fields. Thus, we have
$$\frac{\delta H_{h,n}}{\delta f(p)}=\frac{1}{n!}h'(f(p))\la(p)^n-\frac{1}{(n-1)!}\int\frac{h(f(q))\la(q)^{n-1}}{p-q}dq,$$
and the corresponding vector field is defined as
$$X^{(p)}_{h,n}=\int \frac{\delta H_{h,n}}{\delta f(q)}\,\,\eta^{(q,p)}dq.$$
Explicitly, we have
\beq\label{phvf}
X^{(p)}_{h,n}=\left(\frac{1}{(n-1)!}\int\frac{h(f(q))\la(q)^{n-1}}{p-q}dq-\frac{1}{n!}h'(f(p))\la(p)^n\right)f'(p),
\eeq
and therefore, by construction, these vector fields belong to $\mathcal{V}$.
\begin{thm}
The vector fields \eqref{phvf} satisfy the recurrence relations
\beq\label{dkprecrel}
\nabla_{(q)} X^{(p)}_{h, n+1}=\int c\tbinom{p}{qr}X^{(r)}_{h,n}, \qquad n\in\mathbb{N},
\eeq
of the principal hierarchy. The corresponding commuting flows
\beq\label{dkpphflows}
\de_{t_{h,n}}f(p)=\int V_{h,n}\tbinom{p}{q}f_x(q)dq,\qquad V_{h,n}\tbinom{p}{q}:=\int c\tbinom{p}{qr}X^{(r)}_{h,n},
\eeq
are Hamiltonian of the form
$$\de_{t_{h,n}}f(p)=\left\{f,\mathcal{H}_{h,n+1}\right\}_{LP},$$
where the bracket is the Lie--Poisson bracket \eqref{lpb}, and the Hamiltonian is
$$\mathcal{H}_{h,n+1}=\int H_{h,n+1}dx,$$
with $H_{h,n+1}$ given by \eqref{princhamden}.
\end{thm}
\pf
We first evauate $V_{h,n}$ and find
\begin{align*}
V_{h,n}\tbinom{p}{q}=&\,\,\frac{f'(p)}{p-q}\left(\frac{1}{(n-1)!}\int\frac{h(f(s))\la(s)^{n-1}}{p-s}ds-h'(f(p))\frac{\la(p)^n}{n!}\right)\\
&-\frac{f'(p)}{p-q}\left(\frac{1}{(n-1)!}\int\frac{h(f(s))\la(s)^{n-1}}{q-s}ds-h'(f(q))\frac{\la(q)^n}{n!}\right)\\
&+\delta(p-q)\frac{1}{(n-1)!}\int\frac{f'(r)}{p-r}\int\frac{h(f(s))\la(s)^{n-1}}{r-s}ds\,dr\\
&-\delta(p-q)\frac{1}{n!}\int\frac{f'(r)h'(f(r))\la(r)^n}{p-r}\\
&-\delta(p-q)\left(\frac{\la'(p)}{(n-1)!}\int\frac{h(f(s))\la(s)^{n-1}}{p-s}ds-h'(f(p))\frac{\la(p)^n}{n!}\la'(p)\right).
\end{align*}
On the other hand, we have
\begin{align}
\nabla_{(q)} X^{(p)}_{h, n+1}=&\frac{\delta X^{(p)}_{h, n+1}}{\delta f(q)}+\int\Gamma\tbinom{p}{q\,r}X^{(r)}_{h, n+1}dr\notag\\
=&-\frac{\delta}{\delta f(q)}\left(\frac{\delta H_{h,n+1}}{\delta f(p)}f'(p)\right)-\delta'(q-p)\frac{\delta H_{h,n+1}}{\delta f(q)}\notag\\
=&-f'(p)\frac{\delta^2H_{h,n+1}}{\delta f(q)\delta f(p)}+\frac{\de}{\de p}\left(\frac{\delta H_{h,n+1}}{\delta f(p)}\right)\delta(p-q).\label{ibcompham}
\end{align}
Computing  the second variational derivative, we get
\begin{align*}
\frac{\delta^2H_{h,n+1}}{\delta f(q)\delta f(p)}=& \frac{1}{(n+1)!}\,h''(f(p))\delta(p-q)\la(p)^{n+1}+\frac{1}{n!}h'(f(p))\frac{\la(p)^n}{p-q}\\
&-\frac{1}{n!}h'(f(q))\frac{\la(q)^n}{p-q}-\frac{1}{(n-1)!}\int\frac{h(f(r))\la(r)^{n-1}}{(p-r)(r-q)}dr\\
&+\frac{\pi^2}{(n-1)!}h(f(p))\la(p)^{n-1}\delta(p-q),
\end{align*}
where the last term appears due to the identity \eqref{tricomi} for principal value integrals. On the other hand, the second term in \eqref{ibcompham} gives
\begin{align*}
\frac{\de}{\de p}\frac{\delta H_{h,n+1}}{\delta f(p)}=&\frac{1}{(n+1)!}h''(f(p))f'(p)\la(p)^{n+1}+\frac{1}{n!}h'(f(p))\la(p)^n\la'(p)\\
&-\frac{1}{n!}\int\frac{h'(f(s))f'(s)\la(s)^n}{p-s}ds-\frac{1}{(n-1)!}\int\frac{h(f(s))\la(s)^{n-1}\la'(s)}{p-s}ds.
\end{align*}
Substituting back into \eqref{dkprecrel}, we have that this is satisfied by the vector fields \eqref{phvf}, provided the following identity
\begin{gather*}
\int\frac{h(f(s))\la(s)^{n-1}\la'(s)}{p-s}ds+\pi^2h(f(p))\la(p)^{n-1}f'(p)\\
=\la'(p)\int\frac{h(f(s))\la(s)^{n-1}}{p-s}ds-\int\frac{f'(r)}{p-r}\int\frac{h(f(s))\la(s)^{n-1}}{r-s}ds\,dr,
\end{gather*}
holds. However, this is exactly formula \eqref{tricomi} for $\Phi_1=f'$ and $\Phi_2=h(f)\la^{n-1}$, and therefore the vector fields \eqref{phvf} satisfy the recursion relation \eqref{dkprecrel}. It is now straightforward --using for example \eqref{ibcompham}-- to find the Hamiltonian form of the flows \eqref{dkpphflows}. Indeed, we have
$$\de_{t_{h,n}}f(p)=\!\int\!\nabla_{(q)} X^{(p)}_{h, n+1}f_x(q)dq=-\left\{f(p),\frac{\delta H_{h,n+1}}{\delta f(p)}\right\}_{x,p}=\left\{f(p),\mathcal{H}_{h,n+1}\right\}_{LP},$$
and the theorem is proved.
\epf
\begin{rmk}
By choosing $h(f)=f$, that is, considering the functionals
$$\tilde{H}_{n}=\frac{1}{n!}\int f(q)\la(q)^ndq,$$
we get the classical flows of the dKP hierarchy. For instance, for $n=0$ we have
$$\tilde{H}_{0}=\int f(q)dq=A^0,$$
which is the first of the classical conserved densities. Moreover,
$$\tilde{H}_{1}=\int f(q)\la(q)dq=\int f(q)\Big(q-\pi \text{Hilb}_q[f]\Big) dq=A^1,$$
where we used the fact that a function belonging to $L^2\!\left(\mathbb{R}\right)$ and its Hilbert transform are orthogonal. By using similar identities, one can also prove that
\begin{align*}
\tilde{H}_{2}&=\frac{1}{2}\int f(q)\la(q)^2dq=\frac{1}{2}\int f(q)\Big(q-\pi \text{Hilb}_q[f]\Big)^2 dq\\
&=\frac{1}{2}A^2+\frac{1}{2}\left(A^0\right)^2+\frac{\pi^2}{6}\int f(p)^3dp,
\end{align*}
which differs from the classical conserved density by the last factor, which is a Casimir of the Lie--Poisson bracket.
\end{rmk}
In the above construction of the pricipal hierarchy, we implicitly assumed that the functionals \eqref{princhamden} are conserved densities of the dKP equation. The following proposition fills the gap.
\begin{prop}\label{propgencq}
Let $h(\mu,\nu)$ and $k(\mu,\nu)$ be functions of two variables, sufficiently differentiable, and such that the integrals
\beq\label{gencqdkp}
\mathcal{H}=\iint h(f(p,x),\la(p,x))dp\,dx,\qquad \mathcal{K}=\iint k(f(p,x),\la(p,x))dp\,dx,
\eeq
with $\la$ given by \eqref{lahilbf}, converge. Then, $\left\{\mathcal{H},\mathcal{K}\right\}_{LP}=0,$ and therefore the corresponding Hamiltonian flows
$$f_{t}(p)=\left\{f(p),\mathcal{H}\right\}_{LP},\qquad f_{y}(p)=\left\{f(p),\mathcal{K}\right\}_{LP},$$
commute.
\end{prop}
\pf
In order to prove the proposition, it is sufficient to prove that the condition
$$\frac{\delta}{\delta f(q,y)}\left\{\mathcal{H},\mathcal{K}\right\}_{LP}=0$$
is identically satisfied for any admissible $f$ and for any choice of the functionals $\mathcal{H}$ and $\mathcal{K}$ of the form \eqref{gencqdkp}. Equivalently, we can show that the quantity
\beq\label{halfpoipf}
\frac{\delta}{\delta f(q,y)}\iint f(p,x)\de_{p}\left(\frac{\delta \mathcal{H}}{\delta f(p,x)}\right)\de_{x}\left(\frac{\delta \mathcal{K}}{\delta f(p,x)}\right)dp\,dx,
\eeq
is symmetric in $\mathcal{H}$ and $\mathcal{K}$. We have
$$\frac{\delta \mathcal{H}}{\delta f(p,x)}=\de_{1}h(f(p,x),\la(p,x))-\int\frac{\de_{2}h(f(s,x),\la(s,x))}{p-s}ds,$$
with analogous result for $\mathcal{K}$. Substituting back into \eqref{halfpoipf}, expanding and computing the variational derivative, we obtain a (long) expression, which is proved to be symmetric in $\mathcal{H}$ and $\mathcal{K}$ by using the definition \eqref{lahilbf} of $\la$ and the property \eqref{tricomi} of the Hilbert transform.
\epf
Note that the above proposition is a direct result, in the sense that both its statement and the proof do not depend on the construction of the Frobenius manifold. The particular choice
$k(\mu,\nu)=h(\mu)\tfrac{\nu^n}{n!},$
shows that the functionals \eqref{princhamden} are conserved densities of the principal hierarchy, and, consequently, that the flows of the principal hierarchy pairwise commute. Moreover, by the above proposition, we have that the principal hierarchy can be embedded in a bigger family of commuting Hamiltonian flows, with Hamiltonian of the form \eqref{gencqdkp}. If the density function $k(\mu,\nu)$ of the Hamiltonian is analytic with respect to the second argument, then by Taylor expanding we have that the corresponding conserved density $\mathcal{K}$ can be written as a linear combination of densities \eqref{princhamden}.  This fact reminds of the completeness property of the principal hierarchy, proved in \cite{duzh01} and outlined here in Remark \ref{rmkphham}. However, we do not state that all conserved densities of the hierarchy are of the form \eqref{gencqdkp}; the completeness problem remains thus open. 

We consider now the analogue, in this setting, of the hodograph formula \eqref{hodo}. Since the coordinates $f(p)$ used here are not canonical coordinates, we make use of the weaker formula \eqref{strhod}, thus considering vector fields instead of $(1,1)-$tensors. Moreover, for simplicity, we look for solutions of the dKP equation \eqref{dkp} only; generalization to other members of the hierarchy can be determined --as usual-- by adding the corresponding times and vector fields. Therefore, we set $t_{2}=y$, $t_{3}=t$, and we look for a simultaneous solution $f(p,x,y,t)$ of the flows \eqref{vlasovbenney} and \eqref{vlasovrd}. Within these assumptions, the hodograph formula \eqref{strhod} takes the form
\beq\label{hododkp}
\int c\tbinom{p}{q\,r}\left(xf'(r)+y\,rf'(r)+t\left(r^2+2A^0\right)f'(r)+X^{(r)}_{h,n}\right)dr=0,
\eeq
where we have used the unity vector \eqref{evf}, the vectors \eqref{xvf}  and \eqref{yvf} which  correspond to the flows \eqref{vlasovbenney} and \eqref{vlasovrd} respectively, and one of the vector fields $X_{h,n}$ belonging to the principal hierarchy. Since the factor $f'(p)$ appears in every member of \eqref{hododkp}, one can further factorize the above formula, and look for a function $f$ satisfying
$$x+y\,p+t\left(p^2+2A^0\right)=\frac{\delta H_{h,n}}{\delta f(p)},$$
(compare with the discussion at the end of Section \ref{secfrob}). Moreover, due to Proposition \ref{propgencq}, one can extend the hodograph formula \eqref{hododkp} to a larger class of vector fields, of the form
$$X^{(p)}_K=\int \eta^{(p\,q)}\frac{\delta K}{\delta f(q)}dq=-f'(p)\frac{\delta K}{\delta f(p)},\qquad K=\int k(f,\la)\,dp.$$
These vector fields define symmetries of the hierarchy, and the formula
\beq\label{bestsol}
x+y\,p+t\left(p^2+2A^0\right)=\frac{\delta K}{\delta f(p)},
\eeq
seems thus the most convenient in order to look for solutions of \eqref{vlasovbenney}. 
\begin{rmk}
A formula  similar to \eqref{bestsol} appears -- as a result of a completely different approach -- in \cite{masa08} (see also \cite{masa06}), where the variational derivative of the conserved density in \eqref{bestsol} is replaced by some spectral data obtained by solving a vector (nonlinear) Riemann--Hilbert problem. It would therefore be interesting to obtain a relation between these two equations. 
\end{rmk}
We have obtained condition \eqref{bestsol} by considering -- as done in the rest of the paper -- a continuous index analogue of the finite dimensional theory. We now show that \eqref{bestsol} provides solutions of the dKP equation. Indeed, the following result holds\footnote{I am grateful to Paolo Lorenzoni for this observation}: introducing the functional
\beq\label{genfunk}
\mathcal{H}_{K}=\int \left(A^{0}x+A^{1}y+\left(A^{2}+\left(A^{0}\right)^{2}\right)t-K\right) dx,
\eeq
with $K$ given as above, then the hodograph formula \eqref{bestsol} can be written as the extremal condition
$$\frac{\delta\, \mathcal{H}_{K}}{\delta f(p,x)}=0.$$
Due to this fact, we can prove the following
\begin{prop}
A function $f(p,x,t,y)$ satisfying the hodograph formula \eqref{bestsol} is a solution of the Vlasov equations \eqref{vlasovbenney} and \eqref{vlasovrd}.
\end{prop}
\pf
The function $f$ is obtained by the extremal condition for the functional \eqref{genfunk}. In addition, the latter is a constant of motion for the flows \eqref{vlasovbenney}  and \eqref{vlasovrd}, as shown by the following considerations: the quantities $H^{0}\!=\!A^{0}$, $H^{1}\!=\!A^{1}$, $H^{2}\!=\!A^{2}+\left(A^{0}\right)^{2},$ and $K$ are conserved densities for every flow of the principal hierarchy, and in particular for the flows  \eqref{vlasovbenney}  and \eqref{vlasovrd}. Recalling that $H^{0}_{y}=H^{1}_{x}$, $H^{0}_{t}=H^{2}_{x},$ we obtain
\begin{align*}
\de_{y}\mathcal{H}_{K}&=\int\left(H^{0}_{y}\,x+H^{1}_{y}\,y+H^{1}+H^{2}_{y}\,t+K_{y}\right)dx\\
&=\int\left(H^{1}_{x}\,x+H^{1}+\de_{x}\left(\dots\right)\right)dx=\int\de_{x}\left(H^{1}\,x+\dots\right)dx=0,
\end{align*}
and, similarly,
\begin{align*}
\de_{t}\mathcal{H}_{K}&=\int \left(H^{0}_{t}\,x+H^{1}_{t}\,y+H^{2}_{t}t+H^{2}+K_{t}\right)dx\\
&=\int \left(H^{2}_{x}\,x+H^{2}+\de_{x}\left(\dots\right)\right)dx=\int\de_{x}\left(H^{2}\,x+\dots\right)dx=0.
\end{align*}
Consequently, a function $f$ satisfying \eqref{bestsol} is a stationary point of a conserved quantity for \eqref{vlasovbenney}  and \eqref{vlasovrd}, and it is therefore invariant along these flows.
\epf
As a consequence of the above proposition, we have that the hodograph formula, introduced above as a mere counterpart of the finite dimensional case, produces solutions of the dKP hierarchy, although in an implicit form. A more detailed study of formula \eqref{bestsol} and of solutions of the dKP equation will be considered  in a future publication. We finally remark that the representation of the hodograph formula as a variational condition is valid not only for the dKP case, but for any semi-Hamiltonian system of Egorov type.


\section{Special coordinate sets}\label{secspeccoo}
\subsection{Flat coordinates}
An important feature in the theory of finite dimensional Frobenius manifolds is the existence of the so called flat coordinates, in which the first metric \eqref{stmetric} has constant coefficients. In the preceding sections, we have constructed all important objects of the Frobenius manifold of the dKP hierarchy by using the coordinate $f$, which is non-flat. Here --for completeness-- we consider flat coordinates.
However, as it will be clear from their definition, the existence of flat coordinates together with the requirement of having $f$ real, put severe restictions on the admissible class of functions. We therefore consider $f$ as complex valued and define a new set of coordinates 
\beq\label{flatcoo}
\left\{w(\mu)=f^{-1}(\mu)\right\},
\eeq
which obey the following
\begin{lemma}\label{lemfl}
For the set of coordinates \eqref{flatcoo}, we have
\begin{align}
&\frac{\delta f(p)}{\delta w(\mu)}=\delta(p-w(\mu)),\label{jac1}\\
&\frac{\delta w(\mu)}{\delta f(p)}=\delta(\mu-f(p)). \label{jac2}
\end{align}
Moreover, the following identity holds:
\beq\label{iddel1}
\frac{\delta(w(\mu)-w(\nu))}{f'(w(\mu))}=w'(\mu)\delta(w(\mu)-w(\nu))=\delta(\mu-\nu).
\eeq
\end{lemma}
\pf
Let us consider a test function $K(p)$, and take $p=w(\mu)$. We have
\begin{gather*}
\int K(p)\frac{\delta f(p)}{\delta w(\mu)}dp=\int K(w(\nu))\frac{\delta f(w(\nu))}{\delta w(\mu)} w'(\nu)d\nu\\
=\int K(w(\nu))f'(w(\nu))\delta(\nu-\mu) w'(\nu)d\nu=K(w(\mu))\\
=\int K(p)\delta(p-w(\mu))dp,
\end{gather*}
and therefore \eqref{jac1} holds. The proof of \eqref{jac2} is identical. In order to prove \eqref{iddel1}, we consider a test function $\tilde{K}(\mu,\nu)$, set $\mu=f(p)$, $\nu=f(q)$, and compute
\begin{gather*}
\iint \tilde{K}(\mu,\nu) w'(\mu)\delta(w(\mu)-w(\nu)) d\mu\,d\nu\\=\iint \tilde{K}(f(p),f(q)) w'(f(p))\delta(p-q)f'(p)f'(q) dp\,dq\\
=\int \tilde{K}(f(p),f(p))f'(p) dp=\int \tilde{K}(\mu,\mu)d\mu\\
=\iint \tilde{K}(\mu,\nu)\delta(\mu-\nu)d\mu\,d\nu.
\end{gather*}
\epf
\noindent
We prove now that the \eqref{flatcoo} are flat coordinates for the metric \eqref{stmetric}. Moreover, we give in these coordinates the form of the structure constants \eqref{fconst}; all other objects of the Frobenius manifold can be computed in flat coordinates by similar calculations.
\begin{prop}
In the coordinates \eqref{flatcoo}, the metric \eqref{stmetric} takes the form
$$\eta_{(p,q)}[\omega]=-\delta(p-q).$$
Hence \eqref{flatcoo} are flat coordinates for the Frobenius manifold. Moreover, the structure constants \eqref{fconst} become
$$c\tbinom{\mu}{\nu\,\eta}[w]=\frac{\delta(\mu-\eta)}{w(\mu)-w(\nu)}-\frac{\delta(\nu-\eta)}{w(\mu)-w(\nu)}+\frac{\delta(\mu-\nu)}{w(\nu)-w(\eta)}-\la'(w(\mu))\delta(\nu-\eta)\delta(\mu-\nu),$$
where
$$\la'(w(\mu))=1+\dashint\frac{d\epsilon}{w(\mu)-w(\epsilon)}.$$
\end{prop}
\pf
By using the change of coordinate rules
\begin{gather*}
g_{(p,q)}[w]=\iint g_{(\mu,\nu)}[f]\,\frac{\delta f(\mu)}{\delta w(p)}\,\frac{\delta f(\nu)}{\delta w(q)}\, d\mu d\nu, \\
c\tbinom{p}{q\,r}[w]=\iiint c\tbinom{\zeta}{\mu\,\nu}[f]\,\frac{\delta w(p)}{\delta f(\zeta)}\,\frac{\delta f(\mu)}{\delta w(q)}\,\frac{\delta f(\nu)}{\delta w(r)} \,d\zeta d\mu d\nu,
\end{gather*}
and using Lemma \ref{lemfl}, we obtain the thesis.
\epf
We note that, although in the finite dimensional case the use of flat coordinates simplifies considerably the calculations, this is not the same in the infinite dimensional example considered here. Indeed, the use of the coordinates \eqref{flatcoo} involves delta function identities similar to the one appearing in Lemma \ref{lemfl}, which are more difficult to handle than the $f-$picture approach considered in this paper.

\subsection{Canonical coordinates and Legendre transform}
One of the remarkable results of the paper \cite{cadume09} is the determination, under suitable assumptions, of the canonical coordinates of the Frobenius manifold of the $2D$ Toda hierarchy. We follow here their result to prove that a similar construction holds in the dKP case. In particular, it is convenient to consider the geometrical interpretation outlined at the end of Section \ref{section:sch}, in relation with the Riemann-Hilbert problem.  We thus consider the curve \eqref{planecurve}, and apply to it an analogue of the Legendre transform of classical mechanics: we define the function
$$\mathcal{F}(\alpha,p)=-\alpha \pi f(p)+\la(p),$$
and, fixed $\alpha\in\mathbb{R}$, we consider the extremal condition
\beq\label{extreme}
\frac{\de\mathcal{F}}{\de p}=-\alpha \pi f'(p)+\la'(p)=0.
\eeq
There are two distinct cases to be considered:
\begin{enumerate}
\item If $f'(p)\neq 0$, we introduce the function
$$m(p)=\frac{1}{\pi}\frac{\la'(p)}{f'(p)},$$
so that condition \eqref{extreme} can be written as $\alpha=m(p).$  We consider here only points $p$ where the curve $\gamma$ is not self-intersecting, and such that the direction of every tangent vector on the curve uniquely determines the point on the curve. We denote $\kappa(\alpha)=m^{-1}(\alpha)$ the inverse function, which --by construction-- satisfies
$$\frac{\de\mathcal{F}}{\de p}(\alpha,\kappa(\alpha))=0,$$
and define the Legendre transform-type function
\beq\label{concaco}
r(\alpha)=-\alpha \pi f(\kappa(\alpha))+\la(\kappa(\alpha)).
\eeq
\item If $f'(p)=\la'(p)=0$, namely, if $p$ is a stationary point of $\gamma$, then condition \eqref{extreme} is satisfied independently of $\alpha$. For any stationary point $p_j$, $j=1,\dots, m$, we then define the function
\beq\label{discaco}
r_j=-\alpha \pi f(p_j)+\la(p_j),
\eeq
where $\alpha$ can be chosen to be any fixed real (or possibily complex) number.
\end{enumerate}
\begin{rmk}
The case of reductions of the dispersionless KP equation \cite{gits96} can be characterized by the existence of canonical coordinates of type \eqref{discaco} only.
Moreover, by choosing in \eqref{discaco} $\alpha=i$, we can take the canonical coordinates to be the critical values of the conformal map, that is, the tip of the slits defining the solutions of the system of Loewner equations \cite{gits99}.
\end{rmk}
\begin{prop}
The set of data \eqref{concaco}, \eqref{discaco},  are canonical coordinates for the Frobenius manifold of the dKP hierarchy.
\end{prop}
\pf
We give the proof only for functions of type \eqref{concaco}, the case with singular points can be treated in a similar way. In analogy with the finite dimensional case, $r(\alpha)$ are canonical coordinates if the structure constants \eqref{fconst} take in these coordinates the form
$$c\binom{\alpha}{\beta\,\gamma}[r]=\delta(\alpha-\beta)\delta(\alpha-\gamma).$$
Therefore, we require the coordinate change
$$\int c\tbinom{p}{q\,s}[f]\frac{\delta r(\alpha)}{\delta f(p)}dp=\iint \delta(\alpha-\beta)\delta(\alpha-\gamma)\frac{\delta r(\beta)}{\delta f(q)}\frac{\delta r(\gamma)}{\delta f(s)} d\beta d\gamma,$$
to hold for $r(\alpha)$ given by \eqref{concaco}, and we have
$$\frac{1}{s-q}\left(\frac{\delta r(\alpha)}{\delta f(s)}-\frac{\delta r(\alpha)}{\delta f(q)}\right)-\frac{\delta(q-s)}{f'(q)}\left(\int\frac{f'(p)}{p-q}\frac{\delta r(\alpha)}{\delta f(p)}dp-\la'(q)\frac{\delta r(\alpha)}{\delta f(q)}\right)=\frac{\delta r(\alpha)}{\delta f(q)}\frac{\delta r(\alpha)}{\delta f(s)}.$$
Computing the Jacobian
$$\frac{\delta r(\alpha)}{\delta f(p)}=-\alpha\pi\delta(p-\kappa(\alpha))+\frac{1}{\kappa(\alpha)-p},$$
and substituting back, one gets to the condition
$$\delta(q-s)\left(-\alpha\pi\delta(\kappa(\alpha)-q)+\frac{1}{\kappa(\alpha)-q}\right)\left(-\alpha\pi f'(\kappa(\alpha))+\la'(\kappa(\alpha))\right)=0,$$
which is satisfied due to \eqref{concaco}.
\epf

\section*{Acknowlegments} 
 
I would like to thank John Gibbons for many helpful discussions. I also thank Boris Dubrovin, Tamara Grava, Paolo Lorenzoni, Davide Masoero, Antonio Moro and Marco Pedroni for useful comments and constructive remarks.

\bibliographystyle{plain}
\bibliography{biblio}

\def\cprime{$'$} \def\cprime{$'$} \def\cprime{$'$} \def\cprime{$'$}
  \def\cprime{$'$} \def\cprime{$'$} \def\cprime{$'$} \def\cprime{$'$}
  \def\cprime{$'$} \def\cprime{$'$} \def\cydot{\leavevmode\raise.4ex\hbox{.}}
  \def\cprime{$'$}
\begin{thebibliography}{10}

\bibitem{be73}
D.J. Benney.
\newblock {S}ome properties of long nonlinear waves.
\newblock {\em Stud. Appl. Math.}, 52:45--50, 1973.

\bibitem{bmrwz01}
A.~Boyarsky, A.~Marshakov, O.~Ruchayskiy, P.~Wiegmann, and A.~Zabrodin.
\newblock Associativity equations in dispersionless integrable hierarchies.
\newblock {\em Phys. Lett. B}, 515(3-4):483--492, 2001.

\bibitem{cadume09}
G.~Carlet, B.A. Dubrovin, and L.~Ph. Mertens.
\newblock {I}nfinite-dimensional {F}robenius manifolds for $2+1$ integrable
  systems.
\newblock {\em Math. Ann. {DOI 10.1007/s00208-010-0509-3}}, 2010.

\bibitem{de99}
P.~A. Deift.
\newblock {\em Orthogonal polynomials and random matrices: a
  {R}iemann-{H}ilbert approach}, volume~3 of {\em Courant Lecture Notes in
  Mathematics}.
\newblock New York University, New York, 1999.

\bibitem{dvv91}
R.~Dijkgraaf, E.~Verlinde, and H.~Verlinde.
\newblock {T}opological strings in $d < 1$.
\newblock {\em Nucl. Phys. B}, 352:59, 1991.

\bibitem{do93}
I.~Ya. Dorfman.
\newblock {\em Dirac structures and integrability of nonlinear evolution
  equations}.
\newblock Nonlinear Science: Theory and Applications. John Wiley \& Sons Ltd.,
  Chichester, 1993.

\bibitem{du92}
B.A. Dubrovin.
\newblock {I}ntegrable systems in topological field theory.
\newblock {\em Nucl. Phys. B}, 379:627--689., 1992.

\bibitem{du96}
B.A. Dubrovin.
\newblock Geometry of {$2$}{D} topological field theories.
\newblock In {\em Integrable systems and quantum groups ({M}ontecatini {T}erme,
  1993)}, volume 1620 of {\em Lecture Notes in Math.}, pages 120--348.
  Springer, Berlin, 1996.

\bibitem{du97}
B.A. Dubrovin.
\newblock Flat pencils of metrics and {F}robenius manifolds.
\newblock In {\em Integrable systems and algebraic geometry ({K}obe/{K}yoto,
  1997)}, pages 47--72. World Sci. Publ., River Edge, NJ, 1998.

\bibitem{duno83}
B.A. Dubrovin and S.P. Novikov.
\newblock {H}amiltonian formalism of one-dimensional systems of the
  hydrodynamic type and the {B}ogolyubov-{W}hitham averaging method.
\newblock {\em Dokl. Akad. Nauk SSSR}, 270(4):781--785, 1983.

\bibitem{duzh01}
B.A. Dubrovin and Y.~Zhang.
\newblock {N}ormal forms of hierarchies of integrable {PDE}s, {F}robenius
  manifolds and {G}romov - {W}itten invariants.
\newblock arXiv:math/0108160v1.

\bibitem{duzh06}
B.A. Dubrovin and Y.~Zhang.
\newblock {O}n {H}amiltonian perturbations of hyperbolic systems of
  conservation laws i: {Q}uasi-{T}riviality of bi-{H}amiltonian perturbations.
\newblock {\em Comm. Pure Appl. Math.}, 59(4):559--615, 2006.

\bibitem{elfe05}
P.~Elbau and G.~Felder.
\newblock {D}ensity of eigenvalues of random normal matrices.
\newblock {\em Comm. Math. Phys.}, 259:433---450, 2005.

\bibitem{gesh64}
I.~M. Gel'fand and G.~E. Shilov.
\newblock {\em Generalized functions. {V}ol. {I}: {P}roperties and operations}.
\newblock Academic Press, New York, 1964.

\bibitem{gi81}
J.~Gibbons.
\newblock Collisionless {B}oltzmann equations and integrable moment equations.
\newblock {\em Phys. D}, 3(3):503--511, 1981.

\bibitem{giko94}
J.~Gibbons and Y.~Kodama.
\newblock Solving dispersionless {L}ax equations.
\newblock In {\em Singular limits of dispersive waves (Lyon, 1991)}, volume 320
  of {\em NATO Adv. Sci. Inst. Ser. B Phys.}, pages 61--66. Plenum, New York,
  1994.

\bibitem{gibrai07}
J.~Gibbons and A.~Raimondo.
\newblock Differential geometry of {H}ydrodynamic {V}lasov equations.
\newblock {\em J. Geom. Phys.}, 57(9):1815--1828, 2007.

\bibitem{gits96}
J.~Gibbons and S.P. Tsarev.
\newblock Reductions of the {B}enney equations.
\newblock {\em Phys. Lett. A}, 211(1):19--24, 1996.

\bibitem{gits99}
J.~Gibbons and S.P. Tsarev.
\newblock Conformal maps and reductions of the {B}enney equations.
\newblock {\em Phys. Lett. A}, 258(4-6):263--271, 1999.

\bibitem{kp70}
B.B. {K}adomtsev and V.I. {P}etviashvili.
\newblock {O}n the stability of solitary waves in weakly dispersing media.
\newblock {\em {S}ov. {P}hys. {D}okl.}, 15:539--541, 1970.

\bibitem{ku84}
B.~A. Kupershmidt.
\newblock Normal and universal forms in integrable hydrodynamical systems.
\newblock In {\em Proceedings of the Berkeley-Ames conference on nonlinear
  problems in control and fluid dynamics (Berkeley, Calif., 1983)}, Lie Groups:
  Hist., Frontiers and Appl. Ser. B: Systems Inform. Control, II, pages
  357--378, 1984.

\bibitem{kuma77}
B.A. Kupershmidt and Yu.I. Manin.
\newblock Long wave equations with a free surface. {I}. {C}onservation laws and
  solutions.
\newblock {\em Funktsional. Anal. i Prilozhen.}, 11(3):31--42, 1977.

\bibitem{kuma78}
B.A. Kupershmidt and Yu.I. Manin.
\newblock Long wave equations with a free surface. {II}. {T}he {H}amiltonian
  structure and the higher equations.
\newblock {\em Funktsional. Anal. i Prilozhen.}, 12(1):25--37, 1978.

\bibitem{lope09}
P.~Lorenzoni and M.~Pedroni.
\newblock {N}atural connections for semi-{H}amiltonian systems: the case of the
  $\epsilon-$system.
\newblock arXiv:0912.3697.

\bibitem{lopera09}
P.~Lorenzoni, M.~Pedroni, and A.~Raimondo.
\newblock ${F}-$manifolds and integrable systems of hydrodynamic type.
\newblock arXiv:0905.4054, 2009.

\bibitem{masa06}
S.~V. Manakov and P.~M. Santini.
\newblock {T}he {C}auchy problem on the plane for the dispersionless
  {K}adomtsev--{P}etviashvili equation.
\newblock {\em JETP Lett.}, 83:462--466, 2006.

\bibitem{masa07}
S.~V. Manakov and P.~M. Santini.
\newblock A hierarchy of integrable {PDE}s in $2+1$ dimensions associated with
  $2$-dimensional vector fields.
\newblock {\em Theor. Math. Phys.}, 152:1004--1011, 2007.

\bibitem{masa08}
S.~V. Manakov and P.~M. Santini.
\newblock {O}n the solutions of the d{KP} equation: the nonlinear
  {R}iemann-{H}ilbert problem, longtime behaviour, implicit solutions and wave
  breaking.
\newblock {\em J. Phys. A: Math. Theor.}, 41:055204, 2008.

\bibitem{mawe82}
J.~E. Marsden and A.~Weinstein.
\newblock The {H}amiltonian structure of the {M}axwell-{V}lasov equations.
\newblock {\em Phys. D}, 4(3):394--406, 1981/82.

\bibitem{mus53}
N.~I. Muskhelishvili.
\newblock {\em Singular integral equations. {B}oundary problems of function
  theory and their application to mathematical physics}.
\newblock P. Noordhoff N. V., Groningen, 1953.

\bibitem{pan96}
J.~N. Pandey.
\newblock {\em The {H}ilbert transform of {S}chwartz distributions and
  applications}.
\newblock Pure and Applied Mathematics (New York). John Wiley \& Sons Inc., New
  York, 1996.

\bibitem{resi72}
M.~Reed and B.~Simon.
\newblock {\em Methods of modern mathematical physics. {I}. {F}unctional
  analysis}.
\newblock Academic Press, New York, 1972.

\bibitem{ruso77}
O.~V. Rudenko and S.~I. Soluyan.
\newblock {\em Theoretical foundations of nonlinear acoustics}.
\newblock Consultants Bureau, New York, 1977.
\newblock Translated from the Russian by Robert T. Beyer, Studies in Soviet
  Science.

\bibitem{sato97}
E.B. Saff and V.~Totik.
\newblock {\em Logarithmic Potentials with External Fields}.
\newblock Springer-Verlag, Berlin, 1997.

\bibitem{sch66}
L.~Schwartz.
\newblock {\em Th\'eorie des distributions}.
\newblock Hermann, Paris, 1966.

\bibitem{tata95}
K.~Takasaki and T.~Takebe.
\newblock Integrable hierarchies and dispersionless limit.
\newblock {\em Rev. Math. Phys.}, 7(5):743--808, 1995.

\bibitem{tit86}
E.~C. Titchmarsh.
\newblock {\em Introduction to the theory of {F}ourier integrals}.
\newblock Chelsea Publishing Co., New York, third edition, 1986.

\bibitem{ts90}
S.P. Tsar{\"e}v.
\newblock The geometry of {H}amiltonian systems of hydrodynamic type. {T}he
  generalized hodograph method.
\newblock {\em Izv. Akad. Nauk SSSR Ser. Mat.}, 54(5):1048--1068, 1990.

\bibitem{wiza00}
P.B. Wiegmann and A.~Zabrodin.
\newblock {C}onformal maps and integrable hierarchies.
\newblock {\em Comm. Math. Phys.}, 213:523---538, 2000.

\bibitem{wit91}
E.~Witten.
\newblock Two-dimensional gravity and intersection theory on moduli space.
\newblock In {\em Surveys in differential geometry ({C}ambridge, {MA}, 1990)},
  pages 243--310. Lehigh Univ., Bethlehem, PA, 1991.

\bibitem{giyu00}
L.~Yu and J.~Gibbons.
\newblock The initial value problem for reductions of the {B}enney equations.
\newblock {\em Inverse Problems}, 16(3):605--618, 2000.

\bibitem{zakh69}
E.A. Zabolotskaya and R.V. Khokhlov.
\newblock {Q}uasi-plane {W}aves in the {N}onlinear {A}coustics of {C}onfined
  {B}eams.
\newblock {\em Sov. Phys. Acoustic}, 15:35--40, 1969.

\bibitem{zab09}
A.~Zabrodin.
\newblock Growth of fat slits and dispersionless {KP} hierarchy.
\newblock {\em J. Phys. A}, 42(8):085206, 23, 2009.

\bibitem{zak81}
V.~E. Zakharov.
\newblock {O}n the {B}enney equation.
\newblock {\em Physica D}, 3(1-2):193--202, 1981.

\end{thebibliography}

\end{document}